\documentclass[envcountsame,oribibl,fleqn]{llncs}
\pdfoutput=1  

\title{Counter Machines and Distributed Automata%
  \thanks{To appear in the proceedings of \textsc{Automata} 2018 (published by Springer).}}
\subtitle{A Story about Exchanging Space and Time}
\author{%
  Olivier Carton\inst{1}
  \and
  Bruno Guillon\inst{2}
  \and
  Fabian Reiter\inst{3}
}
\institute{%
  IRIF, Universit\'e Paris Diderot, France \\
  \email{olivier.carton@irif.fr},
  \and
  Department of Computer Science, University of Milan, Italy \\
  \email{guillon.bruno+cs@gmail.com},
  \and
  LSV, Universit\'e Paris-Saclay, France \\
  \email{fabian.reiter@gmail.com}
}

\usepackage[utf8]{inputenc}
\usepackage{microtype}
\usepackage{underscore}  

\usepackage{xifthen}
\usepackage{xspace}
\usepackage{mathtools}
\usepackage{amssymb}
\usepackage{stmaryrd}
\usepackage{pifont}
\usepackage{tikz}
\usetikzlibrary{matrix, positioning, calc, fit, decorations.pathreplacing, shapes.multipart, shapes.symbols}
\usepackage{standalone}
\usepackage[hidelinks]{hyperref}
\hypersetup{pdftitle={Counter Machines and Distributed Automata: %
                      A Story about Exchanging Space and Time},
            pdfauthor={Olivier Carton, Bruno Guillon, and Fabian Reiter}}
\usepackage{sty/commands}

\spnewtheorem*{uProposition}{Proposition}{\normalfont\bfseries}{\itshape}
\spnewtheorem*{uExample}{Example}{\itshape}{\normalfont}

\begin{document}

\maketitle

\begin{abstract}
  We prove the equivalence of two classes of counter machines
  and one class of distributed automata.
  Our counter machines operate on finite words,
  which they read from left to right
  while incrementing or decrementing a fixed number of counters.
  The two classes differ in the extra features they offer:
  one allows to copy counter values,
  whereas the other allows to compute
  copyless sums of counters.
  Our distributed automata,
  on the other hand,
  operate on directed path graphs that represent words.
  All nodes of a path
  synchronously execute the same finite-state machine,
  whose state diagram
  must be acyclic except for self-loops,
  and each node receives as input the state of
  its direct predecessor.
  These devices form a subclass of
  linear-time one-way cellular automata.
\end{abstract}

\section{Introduction}

Space and time are the two standard resources
for solving computational problems.
Typically,
the more of these resources
a computing device has at its disposal,
the harder the problems it can solve.
In this paper,
we consider two types of devices
whose usages of space and time
turn out to be dual to each other.

On the one hand,
we look at \emph{counter machines},
which can use a lot of space.
In the way we define them here,
these devices act as language recognizers.
Just like classical finite automata,
they take a finite word as input,
read it once from left to right,
and then decide whether or not to accept that word.
However,
in addition to having a finite-state memory,
such a machine also has a fixed number of counters,
which can store arbitrarily large integer values
(and are initially set to zero).
The machine has read access to those values
up to some fixed threshold.
Whenever it processes a symbol of the input word,
it can deterministically change its internal state
and simultaneously update each counter~$\cmCounter[1]$
to a new value
that is expressed as the sum of values of
several counters $\cmCounter[2]_1, \dots, \cmCounter[2]_n$
and a constant $\cmConstant$.
(Every update consumes an input symbol,
\ie, there are no epsilon transitions.)
Our main concern are two special cases of this model:
\emph{sumless} counter machines,
which can increment, decrement and copy counter values
but not sum them up,
and \emph{copyless} counter machines,
which can compute arbitrary sums
but not use the same counter more than once per update step.
Both of these conditions entail
that counter values can grow
only linearly with the input length,
and, as we will see,
they yield in fact the same expressive power.

On the other hand,
we look at \emph{distributed automata},
which are devices that can use a lot of time.
For our purposes,
they also act as language recognizers,
but their input word is given in form of a directed path graph
whose nodes are labeled with the symbols of the word
(such that the first symbol is on the source~node).
To run a distributed automaton on such a path,
we first place a copy of the automaton on each node
and initialize it to a state
that may depend on the node's label.
Then,
the execution proceeds
in an infinite sequence of synchronous rounds,
where each node determines its next state
as a function of its own current state
and the current state of its incoming neighbor
(\ie, the node to its left).
Altogether,
there are only a finite number of states,
some of which are considered to be accepting.
The automaton acts as a semi-decider
and accepts the input word
precisely if the last node of the path
visits an accepting state at some point in time.
Here,
we are particularly interested in
those distributed automata
whose state diagram does not contain any directed cycles
except for self-loops;
we call them \emph{quasi-acyclic}.
They have the property that
all nodes stop changing their state
after a number of rounds
that is linear in the length of the input word.
Therefore,
if a quasi-acyclic automaton accepts a given word,
then it does so in linear time.

To sum up,
we have a sequential model and a distributed model
that consume space and time in opposite ways:
given an input word of length~$n$,
a sumless or copyless counter machine
uses time~$n$ and space linear in~$n$,
whereas a quasi-acyclic distributed automaton
uses space~$n$ and linear time.%
\footnote{We assume that counter machines
  store the values of their counters in unary encoding,
  and we measure the space usage of a distributed automaton
  by the number of nodes.}
The purpose of this paper is to show that
there really is a duality between
the space of one model and the time of the other.
In fact,
we will prove that the two models are expressively equivalent.
Besides being of independent interest,
this result also relates to three separate branches of research.

\paragraph{Cellular automata.}
In theoretical computer science,
cellular automata are
one of the oldest and most well-known
models of parallel computation
(see, \eg,~\cite{Kar05}).
They consist of an infinite array
whose cells are each in one of a finite number of states
and evolve synchronously
according to a deterministic local rule.
In this regard,
a distributed automaton
over a labeled directed path\xspace%
can be viewed as
a \emph{(one-dimensional) one-way cellular automaton}
with some permanent boundary symbol
delimiting the input word~\cite{Dyer80}.
This model has been studied as language recognizer,
and differences between real time
(\ie, time~$n$ for an input of length~$n$)
and linear time
(\ie, time in~$\mathcal O(n)$)
have been highlighted
--
see~\cite{Ter12}
for a survey on language recognition by cellular automata.
As explained below,
our work initially takes its motivation
from distributed computing,
hence the choice of ``distributed automata''
rather than ``cellular automata''.
Nevertheless,
the results presented here
may be viewed in terms of languages
recognized by one-way cellular automata,
with the technical difference
that the input words are reversed
with respect to the usual definition of one-way cellular automata.%
\footnote{%
  Contrary to distributed automata,
  one-way cellular automata are usually represented
  with information transiting from right to left,
  that is, a cell receives the state from its right neighbor
  and the leftmost cell decides acceptance of the input,
  see, \eg,~\cite{Kut08}.%
}%

A long-standing open problem
in this area
is the question whether or not
one-way cellular automata
working in unrestricted time
can recognize every language
in~$\textsc{DSpace}(n)$,
\ie, the class of languages
accepted by deterministic Turing Machines
working in linear space.
The latter actually coincides
with the class of languages
accepted by \emph{(two-way) cellular automata}
(see, \eg,~\cite{Kut08}).
By relating a subclass of one-way cellular automata
with counter machines
working in linear space,
our results
might be considered as a new approach
towards describing the expressiveness
of one-way cellular automata.
Our contribution
concerns a class of (reversed) languages
included in the class of languages
recognized by linear-time one-way cellular automata.
Indeed,
the quasi-acyclic restriction on distributed automata
corresponds to a special case
of one-way cellular automata
in which each cell may change its state
only a bounded number of times
during an execution~\cite{Vol81}.
This is
a strict subcase
of one-way cellular automata working in linear time,
as can be deduced,
for instance,
from~\cite[Prop~3]{Vol82}.
More precisely,
quasi-acyclic distributed automata
correspond to
\emph{freezing cellular automata},
which are
cellular automata
in which
each state change of a cell
is increasing according to
some fixed order on the states~\cite{GOT15}.
In particular,
freezing cellular automata
have \emph{bounded communication}~\cite{KM10}.
Conversely, as observed in~\cite{GOT15},
each one-way cellular automaton
with bounded communication
can be easily transformed into an equivalent freezing one.%

\paragraph{Counter machines.}
A classical result due to Minsky states that
Turing machines have the same computational power as
finite-state machines
equipped with two integer counters
that can be arbitrarily often
incremented, decremented, and tested for zero
\cite{Min61}.
Such devices are often referred to as \emph{Minsky~machines}.
Their Turing completeness
led Fischer, Meyer, and Rosenberg
to investigate the space and time complexities of
machines with an arbitrary number of counters,
viewed as language recognizers.
In~\cite{FMR68},
they paid particular attention to \mbox{\emph{real-time machines}},
where the number of increments and decrements per counter
is limited by the length of the input word.
Among many other things,
they showed that increasing the number of counters
strictly increases the expressive power of real-time machines,
and that those devices become even more powerful
if we equip them with the additional ability
to reset counters to zero (in a single operation).
Over four decades later,
Petersen proved in~\cite{Pet11}
that for machines with a single counter,
real time with reset is equivalent to linear time without reset,
and that for machines with at least two counters,
linear time is strictly more expressive.
A further natural extension of real-time machines
is to allow values to be copied from one counter to another
(again, in a single operation).
In~\cite{Dym79},
Dymond showed that
real-time machines with copy
can be simulated by linear-time machines without copy.

The general version of the counter machines defined in this paper
can also be seen as an extension of the real-time machines
of Fischer, Meyer, and Rosenberg.
In addition to the reset and copy operations,
we allow counter values to be summed up.
Our formal notation takes inspiration from
\emph{cost register automata},
which were introduced by Alur~et~al.\
in~\cite{ADDRY13}.
Moreover,
the concept of copylessness is borrowed from there.
The authors are not aware of any previous work
dealing with the specific counter machines defined in this paper.
However,
it follows from \cite[Thm~2.1]{Dym79} and our main result
that sumless and copyless counter machines
form a subclass of the linear-time counter machines
defined in~\cite{FMR68}.

\paragraph{Distributed computing and logic.}
The original motivation for this paper
comes from a relatively recent project
that aims to develop
a form of descriptive complexity \cite{Imm99}
for distributed computing \cite{Lyn96,Pel00}.
In that context,
distributed automata are regarded as
a class of weak distributed algorithms,
for which it is comparatively easy
to obtain characterizations by logical formulas.
Basically,
these automata are the same as those described above,
except that they can run on arbitrary directed graphs
instead of being confined to directed paths.
In order to make this possible,
each node is allowed to see
the set of states of its incoming neighbors
(without multiplicity)
instead of just the state of its left neighbor.
On graphs with multiple edge relations
($\EdgeSet_1, \dots, \EdgeSet_r$),
the nodes see a separate set for each relation.
The first result in this direction
was obtained by Hella~et~al.\
in~\cite{HJKLLLSV15},
where they showed that
distributed automata with constant running time
are equivalent to
a variant of basic modal logic on graphs.
The link with logic was further strengthened
by Kuusisto in~\cite{Kuu13},
where a logical characterization
of unrestricted distributed automata was given
in terms of a modal-logic-based variant of Datalog.
Then,
Reiter showed in~\cite{Rei17} that
the least fixpoint fragment of the modal $\mu$-calculus
captures
an asynchronous variant of quasi-acyclic distributed automata.
Motivated by these connections to modal logic,
a field at the frontier between decidability and undecidability,
the emptiness problem for distributed automata
was investigated in~\cite{KR17}.
The authors observed that the problem is undecidable
for arbitrary automata on directed paths
(which implies undecidability on arbitrary graphs),
as well as for quasi-acyclic automata on arbitrary graphs.
But now,
the main result of the present paper
supersedes both of these findings:
since,
by a simple reduction
from the halting problem for Minsky~machines,
the emptiness problem for sumless and copyless counter machines
is undecidable,
we immediately obtain
that the problem is also undecidable for
quasi-acyclic distributed automata on directed paths.
It must, however, be stressed
that such undecidability results
have been known for a long time
within the community of cellular automata.
For instance,
it was shown by Seidel in~\cite{Sei79}
that the emptiness problem for real-time one-way cellular automata
is undecidable (see also~\cite{Mal02}).
This was later strengthened by Kutrib and Malcher,
who proved in~\cite{KM10} that the problem remains undecidable
even if we restrict ourselves to automata with bounded communication.
Thereby they provided an undecidability result
that is stronger than our corollary,
given that quasi-acyclic distributed automata
do not necessarily work in real time.

\paragraph{Outline.}
The remainder of the paper is devoted to proving
our main result:
\begin{theorem}
  \label{thm:main-result}
  The following three classes of devices
  are effectively equivalent.
  \begin{enumerate}
  \item \label{itm:copyless}
    Copyless counter machines on nonempty finite words.
  \item \label{itm:sumless}
    Sumless counter machines on nonempty finite words.
  \item \label{itm:quasi-acyclic}
    Quasi-acyclic distributed automata on pointed directed paths.
  \end{enumerate}
\end{theorem}
All the necessary definitions are introduced
in Section~\ref{sec:preliminaries}.
The statement then follows from several translations
provided in the subsequent sections:
we have
``\ref{itm:copyless}~$\to$~\ref{itm:sumless}''
by Proposition~\ref{prp:copyless-to-sumless}
in Section~\ref{sec:cm-to-cm},
then
``\ref{itm:sumless}~$\to$~\ref{itm:quasi-acyclic}''
by Propositions~\ref{prp:1-access} and~\ref{prp:cm-to-da}
in Sections~\ref{sec:cm-to-cm} and~\ref{sec:cm-to-da},
and finally
``\ref{itm:quasi-acyclic}~$\to$~\ref{itm:copyless}''
by Proposition~\ref{prp:da-to-cm}
in Section~\ref{sec:da-to-cm}.
We conclude with
a detailed summary of these translations in Section~\ref{sec:conclusion}
(see Figure~\ref{fig:translations})
and some perspectives for future work.

\section{Preliminaries}
\label{sec:preliminaries}

We denote the set of nonnegative integers by
$\defd{\Natural} = \set{0,1,2,\dots}$,
the set of positive integers by
$\defd{\Positive} = \Natural \setminus \set{0}$,
and the set of integers by
$\defd{\Integer} = \set{\dots,-1,0,1,\dots}$.
The power set of any set $S$ is written as \defd{$\powerset{S}$}.
Furthermore,
for values $m, n \in \Integer$
such that $m \leq n$,
we define the interval notation
$\defd{\range[m]{n}} \defeq \setbuilder{i \in \Integer}{m \leq i \leq n}$
and the cutoff function $\cut{m}{n}$,
which truncates its input to yield a number between $m$ and $n$.
The latter is formally defined as
$\defd{\cut{m}{n}} \colon \Integer \to \range[m]{n}$
such that
$\cut{m}{n}(i)$ is equal
to $m$ if $i < m$,
to $i$ if $m \leq i \leq n$,
and to $n$ if $i > n$.

Let $\Alphabet$ be a finite set of symbols.
A \defd{word} over $\Alphabet$ is
a finite sequence
$\Word = \Symbol_1 \dots \Symbol_n$
of symbols in $\Alphabet$.
We write~\defd{$\length{\Word}$}
for the length of $\Word$
and~\defd{$\Alphabet^+$}
for the set of all nonempty words over $\Alphabet$.
A~\defd{language} over $\Alphabet$ is a subset of~$\Alphabet^+$.

\begin{example}[running]
  \label{ex:running-lang}
  As a running example,
  we consider the language~$\runnExLang$
  of nonempty words in $\set{a,b,c}^+$
  whose prefixes all have
  at least as many~$a$'s as~$b$'s
  and at least as many $b$'s as $c$'s:
  $\runnExLang=\setbuilder{w}{\text{for every prefix~$u$ of~$w$, }\length u_a\geq\length u_b\geq\length u_c}$,
  where~$\length w_\sigma$ denotes
  the number of~$\sigma$'s in~$w$,
  for~$\sigma\in\Alphabet$.
  For instance,
  the words~$aaabbc$ and~$aabbac$ belong to~\runnExLang,
  whereas the word~$abacac$ does not.
\end{example}

\subsection{Counter machines}
Let $\cmCounterSet$ be a finite set of counter variables
and $\cmAccess$ be a positive integer.
We denote by \defd{$\cmExpressionSet(\cmCounterSet,\cmAccess)$}
the set of \defd{counter expressions}
over $\cmCounterSet$ and $\cmAccess$
generated by the grammar
$\cmExpression \Coloneqq\, \cmCounter + \cmExpression
                  \,\mid\, \cmConstant$\,,
where
$\cmCounter \in \cmCounterSet$ and
$\cmConstant \in \range[-\cmAccess]{\cmAccess}$.
An \defd{update function} for $\cmCounterSet$ given $\cmAccess$
is a map
$\cmUpdateFunc \in \cmExpressionSet(\cmCounterSet,\cmAccess)^\cmCounterSet$
that assigns a counter expression to each counter variable.

\begin{definition}[Counter Machine]
  \label{def:counter-machine}
  A $\cmCounterNum$-\defd{counter machine}
  with \defd{$\cmAccess$-access}
  over the alphabet~$\Alphabet$
  is a tuple
  $\cmMachine = \tuple{\cmStateSet,\cmCounterSet,\cmInitState,\cmTransFunc,\cmAcceptSet}$,
  where
  $\cmStateSet$
  is a finite set of states,
  $\cmCounterSet$
  is a set containing precisely $\cmCounterNum$~distinct counter variables,
  $\cmInitState \in \cmStateSet$
  is an initial state,
  $\cmTransFunc \colon
   \cmStateSet \times \range[-\cmAccess]{\cmAccess}^\cmCounterSet \times \Alphabet
   \to
   \cmStateSet \times \cmExpressionSet(\cmCounterSet,\cmAccess)^\cmCounterSet$
  is a transition function, and
  $\cmAcceptSet \subseteq \cmStateSet$
  is a set of accepting states.
\end{definition}

Such a counter machine “knows” the exact value of each counter
that lies between the thresholds $-\cmAccess$ and $\cmAccess$;
values smaller than $-\cmAccess$ are “seen” as $-\cmAccess$,
and similarly,
values larger than $\cmAccess$ are “seen” as $\cmAccess$.
Furthermore,
it has the ability to add (in~a single operation)
constants between $-\cmAccess$ and $\cmAccess$ to its counters.
The technical details are explained in the following.

Let
$\cmMachine=\tuple{\cmStateSet,\cmCounterSet,\cmInitState,\cmTransFunc,\cmAcceptSet}$
be a counter machine with $\cmAccess$-access
over the alphabet~$\Alphabet$,
and let
$\Word = \Symbol_1 \dots \Symbol_n$
be a word in $\Alphabet^+$.
A \defd{valuation} of $\cmCounterSet$ is a map
$\cmValuation \in \Integer^\cmCounterSet$
that assigns an integer value
to each counter variable $\cmCounter \in \cmCounterSet$.
The initial valuation is
$\cmInitValuation = \setbuilder{\cmCounter \mapsto 0}{\cmCounter \in \cmCounterSet}$.
Any valuation $\cmValuation \in \Integer^\cmCounterSet$
gives rise to an \defd{extended valuation}
$\cmExtendedValuation \in \Integer^{\,\cmExpressionSet(\cmCounterSet,\cmAccess)}$,
which assigns values to counter expressions in the natural way, \ie,
$\cmExtendedValuation(\cmConstant) = \cmConstant$
and
$\cmExtendedValuation(\cmCounter + \cmExpression) =
 \cmValuation(\cmCounter) + \cmExtendedValuation(\cmExpression)$,
for $\cmConstant \in \range[-\cmAccess]{\cmAccess}$
and $\cmCounter \in \cmCounterSet$.
A \defd{memory configuration} of $\cmMachine$ is a tuple
$\cmCompState = \tuple{\cmState, \cmValuation}
 \in \cmStateSet \times \Integer^\cmCounterSet$.
The \defd{run} of~$\cmMachine$ on~$\Word$
is the sequence of memory configurations
$\cmRun = \tuple{\cmCompState_0, \dots, \cmCompState_n}$
such that
$\cmCompState_0 = \tuple{\cmInitState, \cmInitValuation}$,
and if
$\cmCompState_l = \tuple{\cmState, \cmValuation}$
and
$\cmTransFunc(\cmState,\, \cut{-\cmAccess}{+\cmAccess} \circ \cmValuation,\, \Symbol_{l+1})
 = \tuple{\cmState', \cmUpdateFunc}$,
then
$\cmCompState_{l+1} = \tuple{\cmState',\, \cmExtendedValuation \circ \cmUpdateFunc}$.
The machine~$\cmMachine$ \defd{accepts} the word~$\Word$
if it terminates in an accepting state,
\ie, if
$\cmCompState_n \in \cmAcceptSet \times \Integer^\cmCounterSet$.
The \defd{language} of~$\cmMachine$
(or language \defd{recognized} by~$\cmMachine$)
is the set of all words accepted by~$\cmMachine$.

We call an update function
$\cmUpdateFunc \in \cmExpressionSet(\cmCounterSet,\cmAccess)^\cmCounterSet$
\defd{sumless}
if it does not allow sums of multiple counter variables,
\ie, if for all $\cmCounter[1] \in \cmCounterSet$,
the expression $\cmUpdateFunc(\cmCounter[1])$
is either $\cmConstant$ or $\cmCounter[2] + \cmConstant$,
for some $\cmConstant \in \range[-\cmAccess]{\cmAccess}$
and $\cmCounter[2] \in \cmCounterSet$.
Note that such an update function
allows us to copy the value of one counter to several others,
since the same counter variable~$\cmCounter[2]$
may be used in more than one expression $\cmUpdateFunc(\cmCounter[1])$.
On the other hand,
$\cmUpdateFunc$ is \defd{copyless}
if every counter variable~$\cmCounter[2] \in \cmCounterSet$
occurs in at most one expression $\cmUpdateFunc(\cmCounter[1])$,
and at most once in that expression.
(However, sums of distinct variables are allowed.)
By allowing each counter to be used only once per step,
this restriction ensures that
the sum of all counter values can grow at most linearly
with the length of the input word.
A counter machine $\cmMachine$ is called
\defd{sumless} or \defd{copyless}
if its transition function~$\cmTransFunc$
makes use only of
sumless or copyless update functions,
respectively.
As shown in this paper,
the two notions are expressively equivalent.

\begin{example}[running]
  \label{ex:running-cm}
  The language~$L$ from Example~\ref{ex:running-lang}
  is accepted by the sumless and copyless $2$-counter machine~%
  $\cmMachine=\tuple{\set{\cmState,r},\set{\cmCounter[1],\cmCounter[2]},\cmState,\cmTransFunc,\set\cmState}$,
  with~$\cmTransFunc$ defined by:
  \[
    \cmTransFunc\bigl(s,(c_x,c_y),\sigma\bigr)=
    \left\{
      \renewcommand{\arraystretch}{1.2}
      \begin{array}{l@{}l@{\hspace{1ex}}l@{}l@{\hspace{1.5ex}}l@{\hspace{0.75ex}}l}
        \bigl(\cmState,\,
          \set{
            &
            \cmCounter[1]\coloneqq\cmCounter[1]+1,
            &
            \cmCounter[2]\coloneqq\cmCounter[2]
            &
          }
        \bigr)
        &\text{if }
        s=\cmState
        &
        \text{and }
        \sigma=a,
        \\
        \bigl(\cmState,\,
          \set{
            &
            \cmCounter[1]\coloneqq\cmCounter[1]-1,
            &
            \cmCounter[2]\coloneqq\cmCounter[2]+1
            &
          }
        \bigr)
        &\text{if }
        s=\cmState
        \text{, }
        c_x>0
        &
        \text{and }
        \sigma=b,
        \\
        \bigl(\cmState,\,
          \set{
            &
            \cmCounter[1]\coloneqq\cmCounter[1],
            &
            \cmCounter[2]\coloneqq\cmCounter[2]-1
            &
          }
        \bigr)
        &\text{if }
        s=\cmState
        \text{, }
        c_y>0
        &
        \text{and }
        \sigma=c,
        \\
        \bigl(r,\,
          \set{
            &
            \cmCounter[1]\coloneqq\cmCounter[1],
            &
            \cmCounter[2]\coloneqq\cmCounter[2]
            &
          }
        \bigr)
        &\text{otherwise,}
      \end{array}
    \right.
  \]
  where~$c_x$ and~$c_y$
  denote the values~$\cut{-1}{+1}\circ\cmValuation(x)$
  and~$\cut{-1}{+1}\circ\cmValuation(y)$ respectively.
  Intuitively,
  \cmMachine~uses the counter~$\cmCounter[1]$
  to compare
  the number of~$a$'s
  with those of~$b$'s,
  and the counter~$\cmCounter[2]$
  to compare
  the number of~$b$'s
  with those of~$c$'s.
  The counter~\cmCounter[1] (\resp \cmCounter[2])
  is incremented
  each time the letter~$a$ (\resp~$b$) is read
  and it is decremented
  each time the letter~$b$ (\resp~$c$) is read.
  When a counter with value~$0$
  has to be decremented,
  the machine enters the rejecting sink state~$r$.
\end{example}

\subsection{Distributed automata}
Let $\Alphabet$ be a finite set of symbols.
A \defd{$\Alphabet$-labeled directed graph},
abbreviated \defd{\digraph{}},
is a structure
$\Graph = \tuple{\NodeSet, \EdgeSet, \Labeling}$,
where
$\NodeSet$
is a finite nonempty set of nodes,
$\EdgeSet \subseteq \NodeSet \times \NodeSet$
is a set of directed edges, and
$\Labeling \colon \NodeSet \to \Alphabet$
is a labeling function that assigns a symbol of $\Alphabet$ to each node.
Isomorphic \digraph{s} are considered to be equal.
If $\Node$ is a node in $\NodeSet$,
we call the pair~$\tuple{\Graph,\Node}$
a \defd{pointed \digraph{}}
with \defd{distinguished node}~$\Node$.
Moreover,
if $\Node[1]\Node[2]$ is an edge in $\EdgeSet$,
then $\Node[1]$ is called an \defd{incoming neighbor} of~$\Node[2]$.

A \defd{directed path}, or \defd{\dipath{}},
is a \digraph{}
$\Graph = \tuple{\NodeSet, \EdgeSet, \Labeling}$
that has a distinct \defd{last node}~$\LastNode$
such that each node~$\Node$ in~$\NodeSet$
has at most one incoming neighbor
and exactly one way to reach~$\LastNode$
by following the directed edges in $\EdgeSet$.
A \defd{pointed \dipath{}} is a pointed \digraph{}
$\tuple{\Graph,\LastNode}$
that is composed of a \dipath{} and its last node.
We shall identify each word $\Word \in \Alphabet^+$
with the pointed $\Alphabet$-labeled \dipath{}
of length $\length{\Word}$
whose nodes are labeled with the symbols of~$\Word$,
\ie,
the word
$\Symbol_1 \Symbol_2 \dots \Symbol_n$
will be identified with
the pointed \dipath{} \begin{tikzpicture}[line width=0.5pt,baseline={([yshift=-0.5ex]dipath.base)}]
  \matrix[matrix of math nodes,inner sep=0ex,column sep={4.5ex,between origins},
          nodes={draw,circle,anchor=center,inner sep=0ex,minimum size=2.5ex}] (dipath) {
    \scriptstyle\Symbol_1 & \scriptstyle\Symbol_2 & |[rectangle,draw=none]|{\,\dots\,} & \scriptstyle\Symbol_n  \\
  };
  \coordinate (init) at ([xshift=3.5ex,yshift=1.5ex]dipath-1-4);
  \foreach \x [evaluate = \x as \xplusone using int(\x+1)] in {1,...,3} {
    \draw[->] (dipath-1-\x) -- (dipath-1-\xplusone);
  }
  \draw[->,>=stealth] (init) -- (dipath-1-4);
\end{tikzpicture}\!.

We first give a rather general definition of
distributed automata on arbitrary \digraph{s},
and then slightly modify our notation
for the special case of \dipath{s}.

\begin{definition}[Distributed Automaton]
  \label{def:distributed-automaton}
  A (finite) \defd{distributed automaton}
  over $\Alphabet$-labeled \digraph{s}
  is a tuple
  $\daAutomaton = \tuple{\daStateSet,\daInitFunc,\daTransFunc,\daAcceptSet}$,
  where
  $\daStateSet$
  is a finite set of states,
  $\daInitFunc \colon \Alphabet \to \daStateSet$
  is an initialization function,
  $\daTransFunc \colon \daStateSet \times \powerset{\daStateSet} \to \daStateSet$
  is a transition function, and
  $\daAcceptSet \subseteq \daStateSet$
  is a set of accepting states.
\end{definition}

Let
$\daAutomaton = \tuple{\daStateSet,\daInitFunc,\daTransFunc,\daAcceptSet}$
be a distributed automaton
over $\Alphabet$-labeled \digraph{s},
and let
$\Graph = \tuple{\NodeSet, \EdgeSet, \Labeling}$
be a corresponding \digraph{}.
The (synchronous) \defd{run} of $\daAutomaton$ on $\Graph$
is an infinite sequence
$\daRun = \tuple{\daRun_0, \daRun_1, \daRun_2, \dots}$
of maps
$\daRun_\daTime \colon \NodeSet \to \daStateSet$,
called \defd{configurations},
which are defined inductively as follows,
for $\daTime \in \Natural$ and $\Node[2] \in \NodeSet$:
\begin{equation*}
  \daRun_0(\Node[2]) = \daInitFunc(\Labeling(\Node[2]))
  \qquad\text{and}\qquad
  \daRun_{\daTime+1}(\Node[2]) =
  \daTransFunc
      \bigl(\daRun_\daTime(\Node[2]),\,
            \setbuilder{\daRun_\daTime(\Node[1])}{\Node[1]\Node[2] \in \EdgeSet}
      \bigr).
\end{equation*}
For $\Node \in \NodeSet$,
the automaton~$\daAutomaton$ \defd{accepts}
the pointed \digraph{} $\tuple{\Graph,\Node}$
if $\Node$ visits an accepting state at some point
in the run $\daRun$ of $\daAutomaton$ on $\Graph$,
\ie, if there exists $\daTime \in \Natural$
such that $\daRun_\daTime(\Node) \in \daAcceptSet$.

The above definition could be easily extended
to cover $r$-relational \digraph{s},
\ie, \digraph{s} with $r$~edge relations
$\EdgeSet_1, \dots, \EdgeSet_r$,
for some $r \in \Positive$.
It suffices to choose a transition function of the form
$\daTransFunc \colon \daStateSet \times (\powerset{\daStateSet})^r \to \daStateSet$,
thereby allowing the nodes to see a separate set of states
for each of the $r$~relations.
With this,
one could easily simulate
two-way (one-dimensional) or even higher-dimensional cellular automata.
However,
for our purposes, a single edge relation is enough.

A \defd{trace} of a distributed automaton
$\daAutomaton = \tuple{\daStateSet,\daInitFunc,\daTransFunc,\daAcceptSet}$
is a finite nonempty sequence
$\daState_1, \dots, \daState_n$ of states in $\daStateSet$
such that
for $1 \leq i < n$,
we have
$\daState_i \neq \daState_{i+1}$
and
$\daTransFunc(\daState_i,\daNeighborSet_i) = \daState_{i+1}$
for some $\daNeighborSet_i \subseteq \daStateSet$.
We say that $\daAutomaton$ is \defd{quasi-acyclic}
if its set of traces is finite.
In other words,
$\daAutomaton$ is quasi-acyclic
if its state diagram does not contain any directed cycles,
except for self-loops.
In this case,
we will refer to
$\daTraceLength =
 \max\setbuilder{n}{\text{$\daAutomaton$ has a trace of length $n$}}$
as the \defd{maximum trace length} of~$\daAutomaton$.
Furthermore,
a quasi-acyclic automaton~$\daAutomaton$ is said to have
\defd{at most $(\daLoopNumMinusOne + 1)$ loops per trace} if
$(\daLoopNumMinusOne + 1) =
 \max\setbuilder{n}{\text{$\daAutomaton$ has a trace containing $n$ looping states}}$.
Here,
a \defd{looping state} is a state $\daState \in \daStateSet$
such that
$\daTransFunc(\daState,\daNeighborSet) = \daState$
for some $\daNeighborSet \subseteq \daStateSet$.
Notice that
every trace of a quasi-acyclic automaton must end in a looping state,
since transition functions are defined to be total.
(This is why we write “$\daLoopNumMinusOne + 1$”.)

In this paper,
we regard distributed automata as word acceptors,
and thus we restrict their input to \dipath{s}.
Therefore, in our particular context,
a distributed automaton is the same thing as
a (one-dimensional, reversed) one-way cellular automaton
(see, \eg, \cite{Ter12}).
This allows us to simplify our notation:
transition functions will be written as
$\daTransFunc \colon \daExtendedStateSet \times \daStateSet \to \daStateSet$,
where \defd{$\daExtendedStateSet$}
is a shorthand for $\daStateSet \cup \set{\daNoState}$.
A node whose left neighbor's current state is $\daState[1]$
and whose own current state is $\daState[2]$
will transition to the new state
$\daTransFunc(\daState[1], \daState[2])$;
if there is no left neighbor,
$\daState[1]$~has to be replaced by $\daNoState$.
Note that we have reversed the order of $\daState[1]$ and $\daState[2]$
with respect to
their counterparts in Definition~\ref{def:distributed-automaton},
as this seems more natural when restricted to \dipath{s}.
We say that
the \defd{language} of~$\daAutomaton$
(or language \defd{recognized} by~$\daAutomaton$)
is the set of words, seen as pointed \dipath{s},
accepted by~$\daAutomaton$.

As usual,
we say that two devices
(\ie, counter machines or distributed automata)
are \defd{equivalent}
if they recognize the same language.

\begin{example}[running]
  \label{ex:running-da}
  We describe here a distributed automaton~$\daAutomaton$
  that accepts the language~\runnExLang
  from Example~\ref{ex:running-lang},
  regarded as a set of \dipath{s}
  (see Appendix~\ref{app:preliminaries} for a formal specification).
  To this end,
  we first reformulate the property
  that every prefix contains
  at least as many $a$'s as $b$'s
  and at least as many $b$'s as $c$'s:
  it is equivalent to
  the existence of an injective mapping
  from nodes labeled by~$b$
  to nodes labeled by~$a$
  and from nodes labeled by~$c$
  to nodes labeled by~$b$
  such that
  each node can only be mapped to
  some (possibly indirect) predecessor to its left.
  Our automaton~$\daAutomaton$
  implicitly creates such an injective mapping
  by forwarding all $a$'s and $b$'s to the right
  until they are “consumed”
  by matching $b$'s and $c$'s.

  The device uses two tracks
  that may contain the symbols $a$, $b$, or “$-$”,
  \ie, its states are pairs in
  $\set{a,b,-}\times\set{a,b,-}$.
  Initially,
  a node labeled by the letter~$\sigma\in\Alphabet=\set{a,b,c}$
  is in the state~$(x,y)$,
  where
  $x$ is equal
  to~“$-$” if $\sigma = a$,
  to~$a$ if $\sigma = b$, and
  to~$b$ if $\sigma = c$,
  and
  $y$ is equal
  to~“$-$” if $\sigma = c$, and
  to~$\sigma$ otherwise.
  The first track
  is the \emph{expectation track};
  its content indicates
  which letter
  the node should receive from its left neighbor
  in order to eventually accept
  (the special symbol~“$-$” means ``nothing is expected'')%
  .
  The second track
  is the \emph{communication track};
  its content is sent to the node's right neighbor
  (the special symbol~“$-$” means ``nothing is sent'')%
  .
  If a node is expecting a letter~$\sigma$
  and receives~$\sigma$ from its left neighbor,
  then that node switches to the state~$(-,-)$.
  This means that the node is no longer expecting any letter
  and does not transmit anything to its right neighbor
  (since the letter~$\sigma$ has already been “consumed”).
  Additionally,
  \daAutomaton uses two special states~$\bot$ and~$\top$,
  which propagate errors
  and acceptance, respectively.
  When a node enters one of these two states,
  it stays in that state forever.
  An error 
  always propagates to the right neighbor.
  In contrast,
  a node enters state~$\top$
  if it receives an acceptance message from the left
  (\ie, it receives~$\top$ or~$\emptyset$)
  and its expectation has been fulfilled\xspace%
  .

  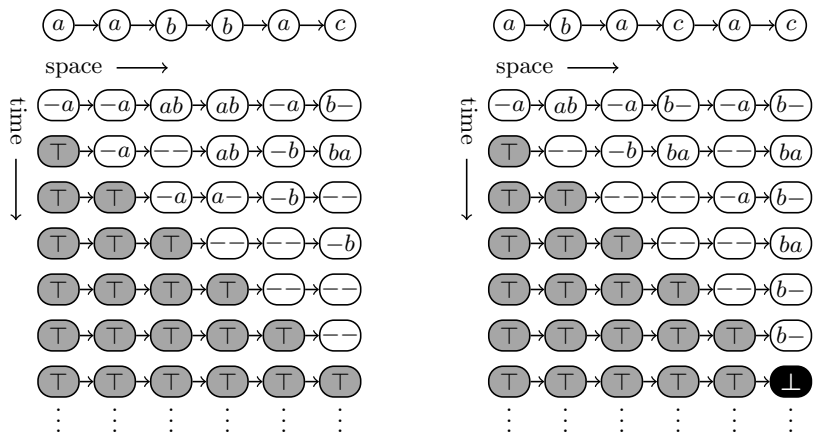
\begin{figure}[tb]
    \centering
    \begin{tikzpicture}[
    semithick,
    every matrix/.style={%
      nodes in empty cells, inner sep=0ex,
      nodes={inner sep=0ex,anchor=center}%
    }
  ]
  \newcommand{\acc}{|[fill=black!35]|\top}
  \newcommand{\rej}{|[fill=black!100,text=white]|\boldsymbol{\bot}}
  \newcommand{\dd}{|[draw=none]|\vdots}
  \begin{scope}
    \matrix[%
      matrix of math nodes,%
      column sep={5.5ex,between origins},%
      nodes={draw,circle,minimum size=3ex,anchor=center}%
    ] (dipath)
    {%
      a & a & b & b & a & c\\
    };
    \foreach \y [count=\yi] in {2,...,6} {
      \draw[->] (dipath-1-\yi) -- (dipath-1-\y);
    }
    \matrix[%
      matrix of math nodes,below=4.5ex of dipath,%
      row sep={4.5ex,between origins},%
      column sep={5.5ex,between origins},%
      nodes={draw,rounded corners=1.4ex,minimum width=4ex,minimum height=3ex,anchor=center}%
    ] (automaton)%
    {%
      -a   & -a   & ab   & ab   & -a   & b-   \\
      \acc & -a   & --   & ab   & -b   & ba   \\
      \acc & \acc & -a   & a-   & -b   & --   \\
      \acc & \acc & \acc & --   & --   & -b   \\
      \acc & \acc & \acc & \acc & --   & --   \\
      \acc & \acc & \acc & \acc & \acc & --   \\
      \acc & \acc & \acc & \acc & \acc & \acc \\[-1.5ex]
      \dd  & \dd  & \dd  & \dd  & \dd  & \dd  \\
    };
    \foreach \x in {1,...,7} {
      \foreach \y [evaluate = \y as \yplusone using int(\y+1)] in {1,...,5} {
        \draw[->] (automaton-\x-\y) -- (automaton-\x-\yplusone);
      }
    }
    \node[above=2ex of automaton.north west,anchor=mid west] (automaton-space) {\small{space}};
    \draw[->] ([xshift=0.5ex]automaton-space.mid east) -- ++(5ex,0);
    \node[left=2ex of automaton.north west,anchor=mid west,rotate=-90] (automaton-time) {\small{time}};
    \draw[->] ([yshift=-0.5ex]automaton-time.mid east) -- ++(0,-6ex);
  \end{scope}
  \begin{scope}[xshift=44ex]
    \matrix[%
      matrix of math nodes,%
      column sep={5.5ex,between origins},%
      nodes={draw,circle,minimum size=3ex,anchor=center}%
    ] (dipath)
    {%
      a & b & a & c & a & c\\
    };
    \foreach \y [count=\yi] in {2,...,6} {
      \draw[->] (dipath-1-\yi) -- (dipath-1-\y);
    }
    \matrix[%
      matrix of math nodes,below=4.5ex of dipath,%
      row sep={4.5ex,between origins},%
      column sep={5.5ex,between origins},%
      nodes={draw,rounded corners=1.4ex,minimum width=4ex,minimum height=3ex,anchor=center}%
    ] (automaton)%
    {%
      -a   & ab   & -a   & b{-} & -a   & b-   \\
      \acc & --   & -b   & ba   & --   & ba   \\
      \acc & \acc & --   & --   & -a   & b-   \\
      \acc & \acc & \acc & --   & --   & ba   \\
      \acc & \acc & \acc & \acc & --   & b-   \\
      \acc & \acc & \acc & \acc & \acc & b-   \\
      \acc & \acc & \acc & \acc & \acc & \rej \\[-1.5ex]
      \dd  & \dd  & \dd  & \dd  & \dd  & \dd  \\
    };
    \foreach \x in {1,...,7} {
      \foreach \y [evaluate = \y as \yplusone using int(\y+1)] in {1,...,5} {
        \draw[->] (automaton-\x-\y) -- (automaton-\x-\yplusone);
      }
    }
    \node[above=2ex of automaton.north west,anchor=mid west] (automaton-space) {\small{space}};
    \draw[->] ([xshift=0.5ex]automaton-space.mid east) -- ++(5ex,0);
    \node[left=2ex of automaton.north west,anchor=mid west,rotate=-90] (automaton-time) {\small{time}};
    \draw[->] ([yshift=-0.5ex]automaton-time.mid east) -- ++(0,-6ex);
  \end{scope}
\end{tikzpicture}
    \caption{The runs of the distributed automaton
      from Example~\ref{ex:running-da} on $aabbac$ and $abacac$.}
    \label{fig:running-da-runs}
  \end{figure}

  Figure~\ref{fig:running-da-runs} shows
  the runs of~\daAutomaton
  on the dipaths~$aabbac$ (accepted)
  and~$abacac$ (rejected).
  Observe that~\daAutomaton
  is not quasi-acyclic,
  since,
  for instance, the last node of the dipath~$aabbac$
  switches from state~$(-,-)$ to~$(-,b)$
  and then again to~$(-,-)$.
\end{example}

\section{Translating between counter machines}
\label{sec:cm-to-cm}

We start with the translation
from copyless to sumless counter machines,
followed by two constructions
that allow us, in some cases,
to focus on counters with non-negative values and $1$-access.
Note that the proofs in this section are merely sketched
(see Appendix~\ref{app:cm-to-cm} for full proofs).

\begin{proposition}
  \label{prp:copyless-to-sumless}
  For every copyless $\cmCounterNum$-counter machine
  with $\cmAccess$-access,
  we can effectively construct an equivalent
  sumless $(2^\cmCounterNum)$-counter machine
  with $(\cmCounterNum \cdot \cmAccess)$-access.
\end{proposition}
\begin{proof}[sketch]
  The idea is simply to introduce a dedicated counter
  for each subset of counters $\cmCounterSubset$
  of the original machine~$\cmMachine$,
  and use this dedicated counter
  to store the sum of values of the counters in~$\cmCounterSubset$.
  Call this sum the \emph{value} of~$\cmCounterSubset$.
  Since $\cmMachine$ is copyless,
  it uses each of its counters at most once
  in any update function~$\cmUpdateFunc$.
  Therefore,
  the next value of $\cmCounterSubset$
  with respect to~$\cmUpdateFunc$
  can be expressed in terms of the current value of
  some other subset $\cmCounterSubset'$
  and a constant between
  $-\card{\cmCounterSubset} \cdot \cmAccess$
  and~$\card{\cmCounterSubset} \cdot \cmAccess$.
  This allows us to derive from~$\cmUpdateFunc$
  a sumless update function~$\cmUpdateFunc'$
  that operates on subsets of counters
  and uses constants in
  $\range[-\cmCounterNum \cmAccess]{\cmCounterNum \cmAccess}$.
  \qed
\end{proof}

Sometimes it is helpful to assume
that a counter machine
never stores any negative values in its counters.
For copyless and sumless machines,
this does not lead to a loss of generality.
We only need the statement for sumless machines,
but in fact Proposition~\ref{prp:copyless-to-sumless} implies
that it also holds for copyless machines
(at the cost of increasing the number of counters).

\begin{proposition}
  \label{prp:nonnegative-counters}
  For every sumless $\cmCounterNum$-counter machine
  with $\cmAccess$-access,
  we can effectively construct an equivalent machine
  that is also sumless with $\cmCounterNum$ counters and $\cmAccess$-access,
  but whose counters never store any negative values.
\end{proposition}
\begin{proof}[sketch]
  It suffices to represent
  each counter~$\cmCounter$ of the original machine
  in such a way that
  the absolute value of~$\cmCounter$ is stored in a counter
  and its sign is retained in finite-state memory.
  As the machine is sumless,
  we do not have to deal with
  the issue of computing
  the sum of a positive and a negative counter~value.
  \qed
\end{proof}

In Definition~\ref{def:counter-machine},
we have introduced counter machines with $\cmAccess$-access,
for some arbitrary $\cmAccess \in \Positive$.
This simplifies some of our proofs,
but we could have imposed $\cmAccess = 1$
without losing any expressive power.
The following proposition states this in full generality,
although the (easier to prove) restriction to sumless machines
would be sufficient to establish our main result.

\begin{proposition}
  \label{prp:1-access}
  For every $\cmCounterNum$-counter machine $\cmMachine$ with $\cmAccess$-access,
  we can effectively construct an equivalent
  $(\cmAccess \cdot \cmCounterNum)$-counter machine $\cmMachine'$ with $1$-access.
  If $\cmMachine$ is copyless or sumless,
  then so is $\cmMachine'$.
  Moreover,
  if $\cmMachine$ is sumless,
  $\cmMachine'$ requires only $\cmCounterNum$ counters.
\end{proposition}
\begin{proof}[sketch]
  The key idea is that $\cmMachine'$ represents
  each counter $\cmCounter$ of $\cmMachine$
  by $\cmAccess$ counters
  $\cmCounter_0, \dots, \cmCounter_{\cmAccess-1}$
  over which the value of $\cmCounter$
  is distributed as uniformly as possible.
  That is,
  the value of $\cmCounter$
  is equal to the sum of the values of
  $\cmCounter_0, \dots, \cmCounter_{\cmAccess-1}$,
  and any two of the latter values differ by at most $1$.
  If~$\cmMachine$ is sumless,
  there is a simpler way:
  it suffices to represent~$\cmCounter$
  by a single counter storing
  the value of~$\cmCounter$ divided by~$\cmAccess$,
  and to keep track of the remainder in the finite-state memory.
  \qed
\end{proof}

\section{From counter machines to distributed automata}
\label{sec:cm-to-da}

Next,
we present the translation from sumless counter machines
to quasi-acyclic distributed automata
(see Appendix~\ref{app:cm-to-da} for a complete proof).

\begin{proposition}
  \label{prp:cm-to-da}
  For every sumless $\cmCounterNum$-counter machine with $1$-access,
  we can effectively construct
  an equivalent quasi-acyclic distributed automaton
  with at most $(\cmCounterNum+2)$ loops per~trace.
\end{proposition}
\begin{proof}[sketch]
  \newcommand*\cmtodaDelay{\ensuremath\top}%
  \newcommand*\cmtodaCount{\ensuremath1}%
  \newcommand*\cmtodaEOCount{\ensuremath0}%
  \newcommand*\cmtodaEOCountbis{\ensuremath\bot}%
  \newcommand*\cmtodaQuiescent{\ensuremath\sharp}%
  \newcommand*\cmtodaVector{\ensuremath y}%
  Our construction uses classical techniques
  from cellular automata theory,
  similar to simulations
  of finite automata
  (see, \eg, \cite[Lem~11]{Kut08})
  and counter machines
  (see, \eg, \cite[Thm~1]{GOT15})
  by one-way cellular automata.
  Let us point out that,
  contrary to the construction in~\cite{GOT15},
  we allow the copy operation on counters here.
  An example of the simulation
  is shown in Figure~\ref{fig:cm2da}.

  \begin{figure}[htb]
    \centering
    \begin{tikzpicture}[
    semithick,
    every matrix/.style = {nodes in empty cells, inner sep=0ex,
                           nodes={inner sep=0ex,anchor=center}},
    brace/.style = {decorate, decoration={brace,mirror,amplitude=0.5ex,raise=0.8ex}},
    brace label/.style = {minimum size=0ex, below=1.3ex},
    pair state/.style={%
        rectangle split,rectangle split parts=2,rectangle split horizontal,%
        rectangle split every empty part={},rectangle split empty part width={0.3em},%
        anchor=center,%
        rounded corners=1.2ex,inner xsep=1pt,inner ysep=0pt,%
        draw=#1%
    },
    pair state/.default={black},
    nonode/.style={draw=none,fill=none,very thin,shape=rectangle,rounded corners=0,inner sep=0,outer sep=0,minimum width=0,anchor=base},
    counter track/.style={nonode,text width=.75em,align=center},
    counter track 1/.style={counter track},
    counter track 2/.style={counter track}
  ]
  \newcommand*\wSymb{\top}
  \newcommand*\cSymb{{\color{white}1}}
  \newcommand*\zSymb{{\color{white}0}}
  \newcommand*\bSymb{\bot}
  \newcommand*\qSymb{\mathbf{\sharp}}
  \newcommand{\inSPair}[3]{%
    |[pair state,#3]|%
    \nodepart{one}\tikz[baseline]\node[counter track 1]{\ensuremath{#1}};%
    \nodepart{two}\tikz[baseline]\node[counter track 2]{\ensuremath{#2}};%
  }
  \newcommand{\inPair}[2]{\inSPair{#1}{#2}{}}%
  \newcommand{\inCPair}[3]{\inSPair{#1}{#2}{rectangle split part fill={#3}}}%
  \newcommand*\ww{\inCPair\wSymb\wSymb{black!35,black!35}}
  \newcommand*\WW{\inPair\wSymb\wSymb}
  \newcommand*\cc{\inCPair\cSymb\cSymb{black,black}}
  \newcommand*\cz{\inCPair\cSymb\zSymb{black,black}}
  \newcommand*\cb{\inCPair\cSymb\bSymb{black,white}}
  \newcommand*\cq{\inCPair\cSymb\qSymb{black,white}}
  \newcommand*\zc{\inCPair\zSymb\cSymb{black,black}}
  \newcommand*\zz{\inCPair\zSymb\zSymb{black,black}}
  \newcommand*\zb{\inCPair\zSymb\bSymb{black,white}}
  \newcommand*\zq{\inCPair\zSymb\qSymb{black,white}}
  \newcommand*\bc{\inPair\bSymb\cSymb}
  \newcommand*\bz{\inCPair\bSymb\zSymb{white,black}}
  \newcommand*\bb{\inPair\bSymb\bSymb}
  \newcommand*\bq{\inPair\bSymb\qSymb}
  \newcommand*\qc{\inPair\qSymb\cSymb}
  \newcommand*\qz{\inCPair\qSymb\zSymb{white,black}}
  \newcommand*\qb{\inPair\qSymb\bSymb}
  \newcommand*\qq{\inPair\qSymb\qSymb}
  \newcommand{\dd}{|[draw=none]|$\vdots$}
  %
  \newcommand{\tip}{\node [draw,fill,signal,signal to=north,signal pointer angle=100,minimum height=0ex,minimum width=4.25ex] {};}
  \newcommand{\x}[1]{\node [draw,minimum height=4.5ex] {$#1$};}
  \newcommand{\sep}{\node [draw,fill,minimum height=0.4ex] {};}
  \newcommand{\ct}[3]{%
    \node (#3) [draw,rectangle split,rectangle split parts=2,rectangle split horizontal,align=center,minimum height=5ex]%
    {\nodepart[text width=2.25ex-\pgflinewidth/2]{one} #1 \nodepart[text width=2.25ex-\pgflinewidth/2]{two} #2};%
  }
  \begin{scope}
  \matrix[matrix of math nodes,column sep={6ex,between origins},
          nodes={draw,minimum height=3ex,minimum width=6ex}] (word) {
    \Symbol[1] & \Symbol[1] & \Symbol[1] & \Symbol[2] & \Symbol[2] & \Symbol[3] \\
  };
  \matrix[matrix of nodes,below=4.5ex of word,row sep={4.5ex,between origins},column sep={6ex,between origins},
          nodes={minimum width=4.5ex,minimum height=4.5ex}] (machine) {
    \tip         & \tip          & \tip         & \tip         & \tip         & \tip         & \tip \\[2.385ex-4.5ex]
    \x{\cmState} & \x{\cmState}  & \x{\cmState} & \x{\cmState} & \x{\cmState} & \x{\cmState} & \x{\cmState} \\[2.45ex-4.5ex]
    \sep         & \sep          & \sep         & \sep         & \sep         & \sep         & \sep \\[2.7ex-4.5ex]
    \ct{0}{0}{A} & \ct{1}{0}{B}  & \ct{2}{0}{C} & \ct{3}{0}{D} & \ct{2}{1}{E} & \ct{1}{2}{F} & \ct{1}{1}{G} \\
  };
  \draw[brace] (G.south west) --
    node[brace label] {$\cmCounter[1]$} ([xshift=-0.1ex]G.south);
  \draw[brace] ([xshift=0.1ex]G.south) --
    node[brace label] {$\cmCounter[2]$} (G.south east);
  \node[above=2ex of machine.north west,anchor=mid west] (machine-time) {\small{time}};
  \draw[->] ([xshift=0.5ex]machine-time.mid east) -- ++(5ex,0);
  \node[left=2ex of machine.north west,anchor=mid west,rotate=-90] (machine-space) {\small{space}};
  \draw[->] ([yshift=-0.5ex]machine-space.mid east) -- ++(0,-4.5ex);
  \node[below=4.2ex of machine.south west,anchor=north west,text width=47ex,align=justify,xshift=-3.5ex] (caption) {%
    The left-hand side illustrates the run of
    the \mbox{$2$-counter} machine from Example~\ref{ex:running-cm}
    on the word~$\Symbol[1]\Symbol[1]\Symbol[1]\Symbol[2]\Symbol[2]\Symbol[3]$.
    On the right-hand side,
    this machine is simulated by a quasi-acyclic distributed automaton
    running on the corresponding
    $\set{\Symbol[1],\Symbol[2],\Symbol[3]}$-labeled dipath.
    Each node of the dipath traverses a sequence of states
    that encodes the memory configuration
    reached by the counter machine
    after reading the node's label.
    (The initial configuration is left implicit.)
    Note that only the two counter tracks
    of the automaton are shown,
    \ie, the transition track is not depicted.
    States represented in gray
    contain the respective node's label
    $\sigma \in \set{\Symbol[1],\Symbol[2],\Symbol[3]}$
    in their transition track,
    whereas black or white states
    contain the machine's state~$\cmState$
    and some counter update function~$\cmUpdateFunc$,
    both determined using the machine's transition function~$\cmTransFunc$.
  };
  \end{scope}
  \begin{scope}
    [xshift=46ex]
    \matrix[%
      matrix of math nodes,%
      column sep={6ex,between origins},%
      nodes={draw,circle,minimum size=3ex,anchor=center}%
    ] (dipath)
    {%
      \Symbol[1] & \Symbol[1] & \Symbol[1] & \Symbol[2] & \Symbol[2] & \Symbol[3] \\
    };
    \foreach \y [count=\yi] in {2,...,6} {
      \draw[->] (dipath-1-\yi) -- (dipath-1-\y);
    }

    \matrix[%
      matrix of nodes,below=4.5ex of dipath,%
      row sep={4.15ex,between origins},%
      column sep={6ex,between origins},%
      nodes={draw,minimum width=4ex,minimum height=3ex,anchor=center}%
    ] (automaton)%
    {%
      \ww & \ww & \ww & \ww & \ww & \ww \\
      \WW & \ww & \ww & \ww & \ww & \ww \\
      \cz & \ww & \ww & \ww & \ww & \ww \\
      \zb & \WW & \ww & \ww & \ww & \ww \\
      \bq & \cz & \ww & \ww & \ww & \ww \\
      \qq & \cb & \WW & \ww & \ww & \ww \\
      \qq & \zq & \cz & \ww & \ww & \ww \\
      \qq & \bq & \cb & \WW & \ww & \ww \\
      \qq & \qq & \cq & \cc & \ww & \ww \\
      \qq & \qq & \zq & \cz & \WW & \ww \\
      \qq & \qq & \bq & \zb & \cc & \ww \\
      \qq & \qq & \qq & \bq & \zc & \WW \\
      \qq & \qq & \qq & \qq & \bz & \cc \\
      \qq & \qq & \qq & \qq & \qb & \zz \\
      \qq & \qq & \qq & \qq & \qq & \bb \\
      \qq & \qq & \qq & \qq & \qq & \qq \\[-1ex]
      \dd & \dd & \dd & \dd & \dd & \dd \\
    };
    \foreach \x in {1,...,16} {
      \foreach \y [evaluate = \y as \yplusone using int(\y+1)] in {1,...,5} {
        \draw[->] (automaton-\x-\y) -- (automaton-\x-\yplusone);
      }
    }
    \node[above=2ex of automaton.north west,anchor=mid west] (automaton-space) {\small{space}};
    \draw[->] ([xshift=0.5ex]automaton-space.mid east) -- ++(5ex,0);
    \node[left=2ex of automaton.north west,anchor=mid west,rotate=-90] (automaton-time) {\small{time}};
    \draw[->] ([yshift=-0.5ex]automaton-time.mid east) -- ++(0,-5.5ex);
  \end{scope}
  \node[above=2.5ex of dipath,anchor=mid,inner sep=0ex] {\small{\emph{Distributed automaton}}};
  \node[above=2.5ex of word,anchor=mid,inner sep=0ex] {\small{\emph{Counter machine}}};
\end{tikzpicture}%
    \caption{%
      Simulating a counter machine
      with a distributed automaton
      to prove Proposition~\ref{prp:cm-to-da}.
      The depicted counter machine is the same as
      in Example~\ref{ex:running-cm}.
      But the resulting automaton
      differs from the one given in Example~\ref{ex:running-da}.
      In particular, it is quasi-acyclic.
    }
    \label{fig:cm2da}
  \end{figure}

  We now explain the main idea.
  On an input \dipath{} corresponding to some word~$\Word$,
  the sequence of states
  traversed by our distributed automaton
  at the $i$-th node
  is an encoding of
  the memory configuration~$\tuple{\cmState,\cmValuation}$
  that is reached by the simulated counter machine
  after reading the $i$-th symbol of~$\Word$.
  (The initial configuration is not encoded.)
  This sequence of states is of the following form:%
  \begin{center}
      \begin{tikzpicture}[semithick,scale=.95,transform shape]
    \newcommand*\CMtoDABlockState[3]{%
      \parbox[c][2ex][c]{2em}{\centering #1}%
      \nodepart{two}%
      \parbox[c][4.5ex][c]{2em}{\centering\tikz[baseline]\node[rotate=90,rectangle,minimum width=2em]{$\cdots$};}%
      \nodepart{three}%
      \parbox[c][2ex][c]{2em}{\centering #2}%
      \nodepart{four}%
      \parbox[c][4ex][c]{2em}{\centering #3}%
    }
    \newcommand*\CMtoDADelay{\ensuremath\top}%
    \newcommand*\CMtoDACount{\ensuremath{\color{white}1}}%
    \newcommand*\CMtoDAEOCount{\ensuremath{\color{white}0}}%
    \newcommand*\CMtoDAEOCountbis{\ensuremath\bot}%
    \newcommand*\CMtoDAQuiescent{\ensuremath\sharp}%
    \newcommand*\CMtoDABlockTrans{%
      {$\cmState,\cmUpdateFunc$}
    }
    \tikzset{%
      mtrack/.style={%
        inner xsep=0,inner ysep=1pt,
        minimum width=2em,
        rectangle split,rectangle split parts=4,
        rectangle split empty part height=4ex,
        rectangle split empty part width=2em,
        draw,text centered%
      },
      GG/.style={rectangle split part fill={black!35,white,black!35,white}},
      B/.style={rectangle split part fill={black,white,white,white}},
      BB/.style={rectangle split part fill={black,white,black,white}}
    }
    \path
      (0,0)        node[mtrack,GG] (delay start)   {\CMtoDABlockState\CMtoDADelay\CMtoDADelay{$\sigma$}}
      ++(0:2.10em) node                            {$\ldots$}
      ++(0:2.10em) node[mtrack,GG] (delay stop)    {\CMtoDABlockState\CMtoDADelay\CMtoDADelay{$\sigma$}}
      ++(0:2.25em) node[mtrack]                    {\CMtoDABlockState\CMtoDADelay\CMtoDADelay\CMtoDABlockTrans}
      ++(0:2.25em) node[mtrack,BB] (counter start) {\CMtoDABlockState\CMtoDACount\CMtoDAEOCount\CMtoDABlockTrans}
      ++(0:2.25em) node[mtrack,B]                  {\CMtoDABlockState\CMtoDACount\CMtoDAEOCountbis\CMtoDABlockTrans}
      ++(0:2.25em) node[mtrack,B]                  {\CMtoDABlockState\CMtoDACount\CMtoDAQuiescent\CMtoDABlockTrans}
      ++(0:2.10em) node                            {$\ldots$}
      ++(0:2.10em) node[mtrack,B]                  {\CMtoDABlockState\CMtoDACount\CMtoDAQuiescent\CMtoDABlockTrans}
      ++(0:2.25em) node[mtrack,B]  (counter stop)  {\CMtoDABlockState\CMtoDAEOCount\CMtoDAQuiescent\CMtoDABlockTrans}
      ++(0:2.25em) node[mtrack]                    {\CMtoDABlockState\CMtoDAEOCountbis\CMtoDAQuiescent\CMtoDABlockTrans}
      ++(0:2.25em) node[mtrack]    (omega start)   {\CMtoDABlockState\CMtoDAQuiescent\CMtoDAQuiescent\CMtoDABlockTrans}
      ++(0:2.25em) node[mtrack]                    {\CMtoDABlockState\CMtoDAQuiescent\CMtoDAQuiescent\CMtoDABlockTrans}
      ++(0:2.10em) node                            {$\ldots$}
      ++(0:1.00em) coordinate[rectangle,minimum height=12.5ex,inner sep=0ex]  (r)
      ;
    \draw[decorate,decoration={brace,amplitude=0.8ex,raise=3.5pt,mirror}]
      (delay start.north west)  --  ($(delay start.south west)+(0,4.1ex)$)
      node[xshift=-6pt,yshift=1pt,text width=4em,midway,left,align=right] {\small $\cmCounterNum$ counter\\[-4pt]tracks}
      ;
    \draw[decorate,decoration={brace,amplitude=0.8ex,raise=3.5pt,mirror}]
      ($(delay start.south west)+(0,3.9ex)$)  --  (delay start.south west)
      node[xshift=-6pt,yshift=1pt,text width=4em,midway,left,align=right] {\small transition\\[-4pt]track}
      ;
    \draw[decorate,decoration={brace,amplitude=0.8ex,raise=3.5pt}]
      (delay start.north west)  --  (delay stop.north east)
      node[yshift=6pt,midway,above]  {\small delay phase}
      ;
    \draw[decorate,decoration={brace,amplitude=0.8ex,raise=3.5pt}]
      (counter start.north west)  --  (counter stop.north east)
      node[yshift=6pt,midway,above]  {\small counter valuation phase}
      ;
    \draw[decorate,decoration={brace,amplitude=0.8ex,raise=3.5pt}]
      (omega start.north west)  --  (r.north |- omega start.north west)
      node[yshift=6pt,midway,above]  {\small final phase}
      ;
    \path
      (delay start.south)
      ++(270:10pt)    edge[->]
      node[fill=white,inner ysep=0,inner xsep=3pt,yshift=1pt] {\small time}
      ++(0:4cm)
      ;
  \end{tikzpicture}
  \end{center}
  Here,
  each rectangular block represents a state of the distributed automaton.
  The symbol~$\sigma$ corresponds to the node's label
  and~$\cmUpdateFunc$ is the update function that has been used
  to enter the memory configuration~$(\cmState,\cmValuation)$.
  Counter values are encoded in unary,
  \ie, the value~$\cmValuation(\cmCounter)$
  of a counter~$\cmCounter$
  is the number of $1$'s on the associated counter track.
  (By Proposition~\ref{prp:nonnegative-counters},
  we assume the values are never negative.)

  The delay phase is used
  to leave enough time
  for information to transit.
  We increase it by~$2$ at each position,
  in order to be able to compute decrementation.
  Hence, at the $i$-th node, the delay phase lasts for~$(2i-1)$ rounds.
  (This corresponds to the gray states in Figure~\ref{fig:cm2da}.)

  Since each counter track
  associated with a counter~$\cmCounter$
  contains a sequence of the form
  $\cmtodaDelay^{2i}1^{\cmValuation(\cmCounter)}\cmtodaEOCount\cmtodaEOCountbis\cmtodaQuiescent^\omega$,
  we are guaranteed that the simulating distributed automaton has at most~$(\cmCounterNum+2)$ loops per trace.
  \qed
\end{proof}

\section{From distributed automata to counter machines}
\label{sec:da-to-cm}

As the last piece of the puzzle,
we now show how to convert a quasi-acyclic distributed automaton
into an equivalent copyless counter machine
(see Appendix~\ref{app:da-to-cm} for a complete proof).

\begin{proposition}
  \label{prp:da-to-cm}
  For every quasi-acyclic distributed automaton
  with at most $(\daLoopNumMinusOne + 1)$ loops per trace
  and maximum trace length~$\daTraceLength$,
  we can effectively construct
  an equivalent copyless $\daLoopNumMinusOne$-counter machine
  with $\daTraceLength$-access.
\end{proposition}
\begin{proof}[sketch]
  Basically,
  after our counter machine~$\cmMachine$ has read
  the $i$-th symbol of the input word~$\Word$,
  its memory configuration will represent
  the sequence of states traversed by
  the simulated distributed automaton~$\daAutomaton$
  at the $i$-th node
  of the \dipath{} corresponding to~$\Word$.
  This exploits the quasi-acyclicity of~$\daAutomaton$
  to represent
  the infinite sequence of states traversed by a node
  as a finite sequence of pairs in
  $\daStateSet \times (\Positive \cup \set{\infty})$,
  where
  values other than $1$ and $\infty$ are stored in the counters.
  An example illustrating the construction
  is provided in Figure~\ref{fig:da-to-cm}.

  \begin{figure}[htb]
    \centering
    \begin{tikzpicture}[
    semithick,
    every matrix/.style = {nodes in empty cells, inner sep=0ex,
                           nodes={inner sep=0ex,anchor=center}},
    stateA/.style = {fill=black!25},
    stateB/.style = {fill=black!65, text=white},
    stateC/.style = {fill=black!15},
    stateD/.style = {fill=black!50, text=white},
    stateE/.style = {fill=black!0},
    stateF/.style = {fill=black!100, text=white},
    historyNode/.style = {draw,align=center,execute at begin node={\setlength{\baselineskip}{2ex}}},
    acceptBox/.style = {draw,minimum height=5.4ex,minimum width=2.4ex},
    brace/.style = {decorate, decoration={brace,raise=1ex}},
    brace label/.style = {minimum size=0ex, right=1.5ex}
  ]
  \newcommand{\sA}{|[stateA]|$\daStateA$}
  \newcommand{\sB}{|[stateB]|$\daStateB$}
  \newcommand{\sC}{|[stateC,double]|$\daStateC$}
  \newcommand{\sD}{|[stateD]|$\daStateD$}
  \newcommand{\sE}{|[stateE]|$\daStateE$}
  \newcommand{\sF}{|[stateF]|$\daStateF$}
  \newcommand{\dd}{|[draw=none]|$\vdots$}
  \newcommand{\tIni}{\node [historyNode] {$\daNoState$ \\ $\infty$};}
  \newcommand{\tA}{\node [historyNode,stateA] {$\daStateA$ \\ $\cmCounter[1]$};}
  \newcommand{\tB}{\node [historyNode,stateB] {$\daStateB$ \\ $1$};}
  \newcommand{\tC}{\node[stateC,acceptBox] {}; \node [historyNode] {$\daStateC$ \\ $1$};}
  \newcommand{\tD}{\node [historyNode,stateD] {$\daStateD$ \\[0.5ex] $\cmCounter[1]$};}
  \newcommand{\tE}{\node [historyNode,stateE] {$\daStateE$ \\ $\cmCounter[2]$};}
  \newcommand{\tF}{\node [historyNode,stateF] {$\daStateF$ \\ $\infty$};}
  \newcommand{\tEinf}{\node [historyNode,stateE] {$\daStateE$ \\ $\infty$};}
  \newcommand{\tip}{\node [draw,fill,signal,signal to=north,signal pointer angle=100,minimum height=0ex,minimum width=2.75ex] {};}
  \newcommand{\x}[1]{\node [draw,minimum height=3ex] {$#1$};}
  \newcommand{\X}[2]{\node (#2) [draw,minimum height=3ex] {$#1$};}
  \newcommand{\sep}{\node [draw,fill,minimum height=0.4ex] {};}
  \matrix[matrix of math nodes,column sep={4.5ex,between origins},
          nodes={draw,circle,minimum size=3ex}] (dipath) {
    \Symbol[2] & \Symbol[1] & \Symbol[2] & \Symbol[2] & \Symbol[1] & \Symbol[1] & \Symbol[1] \\
  };
  \foreach \y [evaluate = \y as \yplusone using int(\y+1)] in {1,...,6} {
    \draw[->] (dipath-1-\y) -- (dipath-1-\yplusone);
  }
  \matrix[matrix of nodes,below=4.5ex of dipath,row sep={4.5ex,between origins},column sep={4.5ex,between origins},
          nodes={draw,circle,minimum size=3ex}] (automaton) {
    \sD & \sA & \sD & \sD & \sA & \sA & \sA \\
    \sE & \sA & \sD & \sE & \sA & \sA & \sA \\
    \sE & \sB & \sD & \sE & \sB & \sA & \sA \\
    \sE & \sC & \sE & \sE & \sC & \sA & \sA \\
    \sE & \sF & \sE & \sE & \sF & \sE & \sA \\
    \sE & \sF & \sF & \sE & \sF & \sF & \sB \\
    \sE & \sF & \sF & \sF & \sF & \sF & \sF \\[-1.5ex]
    \dd & \dd & \dd & \dd & \dd & \dd & \dd \\
  };
  \foreach \x in {1,...,7} {
    \foreach \y [evaluate = \y as \yplusone using int(\y+1)] in {1,...,6} {
      \draw[->] (automaton-\x-\y) -- (automaton-\x-\yplusone);
    }
  }
  \matrix[matrix of math nodes,right=12ex of dipath,column sep={4.5ex,between origins},
          nodes={draw,minimum height=3ex,minimum width=4.5ex}] (word) {
    \Symbol[2] & \Symbol[1] & \Symbol[2] & \Symbol[2] & \Symbol[1] & \Symbol[1] & \Symbol[1] \\
  };
  \matrix[matrix of nodes,below=4.5ex of word,row sep={6ex,between origins},column sep={4.5ex,between origins},
          nodes={minimum width=3ex,minimum height=6ex}] (machine) {
    \tip  & \tip   & \tip  & \tip  & \tip  & \tip  & \tip  & \tip \\[1.63ex-6ex]
    \x{0} & \x{1}  & \x{2} & \x{3} & \x{1} & \x{2} & \x{4} & \X{5}{counter-1} \\[3ex-6ex]
    \x{0} & \x{0}  & \x{0} & \x{2} & \x{5} & \x{0} & \x{1} & \X{0}{counter-2} \\[1.7ex-6ex]
    \sep  & \sep   & \sep  & \sep  & \sep  & \sep  & \sep  & \sep \\[3.2ex-6ex]
    \tIni & \tD    & \tA   & \tD   & \tD   & \tA   & \tA   & \tA \node (first-substate) {}; \\
          & \tEinf & \tB   & \tE   & \tE   & \tB   & \tE   & \tB \\
          &        & \tC   & \tF   & \tF   & \tC   & \tF   & \tF \node (last-substate) {}; \\
          &        & \tF   &       &       & \tF   &       & \\
  };
  \draw[brace] (counter-1.north east) --
    node[brace label] {$\cmCounter[1]$} ([yshift=0.2ex]counter-1.south east);
  \draw[brace] ([yshift=-0.2ex]counter-2.north east) --
    node[brace label] {$\cmCounter[2]$} (counter-2.south east);
  \draw[brace] (first-substate.north east) --
    node[brace label,rotate=-90,anchor=south] {\small{finite state}} (last-substate.south east);
  \node[above=2ex of automaton.north west,anchor=mid west] (automaton-space) {\small{space}};
  \draw[->] ([xshift=0.5ex]automaton-space.mid east) -- ++(5ex,0);
  \node[left=2ex of automaton.north west,anchor=mid west,rotate=-90] (automaton-time) {\small{time}};
  \draw[->] ([yshift=-0.5ex]automaton-time.mid east) -- ++(0,-6ex);
  \node[above=2ex of machine.north west,anchor=mid west] (machine-time) {\small{time}};
  \draw[->] ([xshift=0.5ex]machine-time.mid east) -- ++(5ex,0);
  \node[left=2ex of machine.north west,anchor=mid west,rotate=-90] (machine-space) {\small{space}};
  \draw[->] ([yshift=-0.5ex]machine-space.mid east) -- ++(0,-5ex);
  \node[above=2.5ex of dipath,anchor=mid,inner sep=0ex] {\small{\emph{Distributed automaton}}};
  \node[above=2.5ex of word,anchor=mid,inner sep=0ex] {\small{\emph{Counter machine}}};
\end{tikzpicture}
    \caption{
      Simulating a distributed automaton with a counter machine
      to prove Proposition~\ref{prp:da-to-cm}.
      The left-hand side depicts the run of
      a quasi-acyclic distributed automaton
      on the $\set{\Symbol[1],\Symbol[2]}$-labeled \dipath{}
      that corresponds to the word
      $\Symbol[2]\Symbol[1]\Symbol[2]\Symbol[2]\Symbol[1]\Symbol[1]\Symbol[1]$.
      This automaton has at most three loops per trace;
      its set of states~$\daStateSet$
      consists of the states $\daStateA, \daStateD, \daStateE, \daStateF$,
      which have self-loops,
      and the states $\daStateB, \daStateC$,
      which do not.
      On the right-hand side,
      the automaton is simulated by
      a copyless $2$-counter machine
      whose memory configurations encode
      infinite sequences of states of the automaton
      as finite sequences of pairs in
      $\daStateSet \times (\Positive \cup \set{\infty})$.
      Values different from $1$ and $\infty$ are stored
      in the two counters $\cmCounter[1]$ and $\cmCounter[2]$.
    }
    \label{fig:da-to-cm}
  \end{figure}

  The crux of the proof
  is the following:
  if the $i$-th node remains in the same state
  for more than $\daTraceLength$ rounds,
  then (by quasi-acyclicity)
  the sequence of states traversed
  during that time by the $(i+1)$-th node
  must become constant
  (\ie, repeating always the same state)
  no later than the $\daTraceLength$-th round.
  Thus,
  to compute the entire state sequence
  of the $(i+1)$-th node,
  $\cmMachine$ does not need to know
  the exact numbers of state repetitions
  in the $i$-th node's sequence.
  It only needs to know these numbers
  up to threshold $\daTraceLength$
  and be able to sum them up.
  \qed
\end{proof}

\section{Conclusion}
\label{sec:conclusion}

\begin{figure}[htb]
  \centering
  \begin{tikzpicture}[
   semithick,
   main/.style={draw,rounded corners,align=left,inner sep=1.5ex},
   aux/.style={align=left,font=\scriptsize,inner sep=1ex},
   hv/.style={to path={-| (\tikztotarget)}},
   vh/.style={to path={|- (\tikztotarget)}},
  ]
  \path
    node[main]
    (ccm) {Copyless \\ counter machines}
    (ccm.south) node[aux,below]
    (ccm-from) {$\cmCounterNum$ counters \\ $\cmAccess$-access}
    (ccm.east) node[aux,right]
    (ccm-to) {$\daLoopNumMinusOne$ counters \\ $\,\daTraceLength$-access}
    (ccm.north east) node[anchor=north east]
    {(\ref{itm:copyless})}
    (ccm) ++(270:27ex) node[main]
    (scm) {Sumless \\ counter machines}
    (scm.east) node[aux,right]
    (scm-from) {$\cmCounterNum$ counters \\ $1$-access}
    (scm.south) node[aux,below]
    (scm-loop) {$\cmCounterNum$ counters \\ $\cmAccess$-access}
    (scm.north) node[aux,above]
    (scm-to) {$\;2^\cmCounterNum$ counters \\ $(\cmCounterNum \cdot \cmAccess)$-access}
    (scm.north east) node[anchor=north east]
    {(\ref{itm:sumless})}
    ($(ccm.east)!0.5!(scm.east)$) ++(0:37ex) node[main]
    (qda) {Quasi-acyclic \\ distributed automata}
    (qda.north) node[aux,above]
    (qda-from) {$\leq (\daLoopNumMinusOne+1)$ $\text{loops}/\text{trace}$ \\ max.\ trace length~$\daTraceLength$}
    (qda.south) node[aux,below]
    (qda-to) {$\leq (\cmCounterNum+2)$ $\text{loops}/\text{trace}$}
    (qda.north east) node[anchor=north east]
    {(\ref{itm:quasi-acyclic})}
    (ccm.east) ++(0:32ex) coordinate
    (qda->ccm)
    (scm.east) ++(0:32ex) coordinate
    (scm->qda)
    ;
  \draw[->]
    (ccm-from) --
     node[right] {Prop.~\ref{prp:copyless-to-sumless}}
    (scm-to);
  \draw[->,rounded corners]
    ([xshift=-0.5ex,yshift=-1.5ex]scm-loop.east) -|
     node[near start,above,xshift=0.5ex] {Prop.~\ref{prp:1-access}}
    ([xshift=1.5ex]scm-from.south);
  \draw[->,rounded corners]
    (scm-from.east) -|
     node[pos=0.2,above] {Prop.~\ref{prp:cm-to-da}}
    (qda-to.south);
  \draw[->,rounded corners]
    (qda-from.north) |-
     node[pos=0.8,below] {Prop.~\ref{prp:da-to-cm}}
    (ccm-to);
\end{tikzpicture}
  \caption{
    The translations involved in the proof of Theorem~\ref{thm:main-result}.
  }
  \label{fig:translations}
\end{figure}
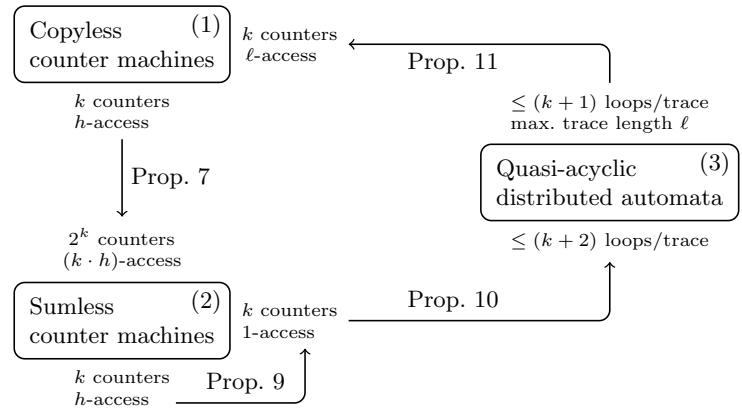

We have now completed the proof of Theorem~\ref{thm:main-result},
which states the equivalence of
(\ref{itm:copyless})~copyless and (\ref{itm:sumless})~sumless
counter machines on finite words
and
(\ref{itm:quasi-acyclic})~quasi-acyclic distributed automata
on pointed \dipath{s}.
More precisely,
we have established the following translatability results,
which are visualized in Figure~\ref{fig:translations}:
\begin{enumerate}
\item \label{itm:copyless-to-sumless}
  A copyless $\cmCounterNum$-counter machine
  with $\cmAccess$-access
  can be translated into
  an equivalent sumless $(2^\cmCounterNum)$-counter machine
  with $(\cmCounterNum \cdot \cmAccess)$-access
  (by Proposition~\ref{prp:copyless-to-sumless}).
\item \label{itm:cm-to-da}
  A sumless $\cmCounterNum$-counter machine with $\cmAccess$-access,
  can be transformed into
  an equivalent (sumless $\cmCounterNum$-counter) machine
  that has merely $1$-access
  (by Proposition~\ref{prp:1-access}),
  which in turn can be translated into
  an equivalent quasi-acyclic distributed automaton
  with at most $(\cmCounterNum+2)$ loops per~trace
  (by Proposition~\ref{prp:cm-to-da}).
\item \label{itm:da-to-cm}
  A quasi-acyclic distributed automaton
  with at most~$(\daLoopNumMinusOne + 1)$ loops per trace
  and maximum trace length~$\daTraceLength$
  can be translated into
  an equivalent copyless $\daLoopNumMinusOne$-counter machine
  with $\daTraceLength$-access
  (by Proposition~\ref{prp:da-to-cm}).
\end{enumerate}

This cycle of translations suggests that
the number of counters of copyless and sumless counter machines
is closely related to
the maximum number of loops per trace of quasi-acyclic distributed automata.
However,
the precise relationship is left open.
In particular, as of the time of writing,
the authors do not know whether
the exponential blow-up of the number of counters
in Proposition~\ref{prp:copyless-to-sumless}
could be avoided.

In addition,
there are several natural directions
in which the present work might be extended.
First of all,
the models of computation concerned by Theorem~\ref{thm:main-result}
are special cases of two more general classes of word acceptors,
namely the unrestricted
counter machines of Definition~\ref{def:counter-machine}
and the unrestricted
distributed automata of Definition~\ref{def:distributed-automaton}
on pointed \dipath{s}
(or equivalently, reversed one-way cellular automata).
It is thus natural to ask
whether our result carries over to
stronger (sub)classes of counter machines and distributed automata.
Instead of counter machines,
one might also consider sequential machines
with more freely accessible memory,
such as restricted read-write tapes.
Second,
one could conversely try to establish
similar connections for weaker classes of devices.
In particular,
it would be interesting
to find a distributed characterization of
the real-time counter machines
of Fischer, Meyer, and Rosenberg~\cite{FMR68},
which are both copyless and sumless.
Third,
all of the models considered in this paper are one-way,
in the sense that
counter machines scan their input from left to right
and distributed automata on \dipath{s} send information from left to right.
Hence,
another obvious research direction
would be to investigate the connections between
(suitably defined) two-way versions.
Finally,
for the sake of presentational simplicity,
we have only looked at deterministic models.
It seems, however, that our proofs could be easily extended
to cover nondeterministic or even alternating devices.
We leave this open for future work.

\subsection*{Acknowledgments}

We are grateful to the anonymous reviewers
for their constructive comments.
We also thank Martin Kutrib and Pierre Guillon
for interesting discussions,
especially concerning the connection of our results
with the field of cellular automata.
This work was partially supported by the
\href{http://www.lsv.fr/~bouyer/equalis/index.php}{ERC project EQualIS (FP7-308087)}
and the
\href{http://delta.labri.fr}{DeLTA project (ANR-16-CE40-0007)}.

\bibliography{cm-paper.bib}

\newpage
\appendix
\section{Supplement to the preliminaries}
\label{app:preliminaries}

\begin{uExample}[\ref{ex:running-da}]
  The distributed automaton~$\daAutomaton$
  from Example~\ref{ex:running-da} in Section~\ref{sec:preliminaries}
  can be formally described as follows:
  $\daAutomaton = \tuple{\daStateSet,\daInitFunc,\daTransFunc,\daAcceptSet}$,
  where
	\begin{itemize}
		\item $\daStateSet=\set{a,b,-}^2\cup\set{\top,\bot}$\xspace%
			\,and\,
			$\daAcceptSet=\set{\top}$;
		\item $\daInitFunc\colon\Alphabet\to\daStateSet$ is defined by
			$\daInitFunc(a)=(-,a)$,\,
			$\daInitFunc(b)=(a,b)$
			and
			$\daInitFunc(c)=(b,-)$;
		\item $\daTransFunc\colon\daExtendedStateSet\times\daStateSet\to\daStateSet$
			is defined by
			\[
				\begin{aligned}
					\begin{aligned}
						\daTransFunc\tuple{(x',y'),(x,y)}&=
						\left\{
							\begin{aligned}
                                                          &(-,-)	&&\text{if~$y'=x$,}			\\
							 &(x,y')	&&\text{otherwise,}				\\
							\end{aligned}
						\right.
						\\
						\left.
						\begin{gathered}
							\daTransFunc\tuple{\top,(x,y)}
							\\
							\daTransFunc\tuple{\emptyset,(x,y)}
						\end{gathered}
						\right]
						&=
						\left\{
							\begin{aligned}
								&\top	&&\text{if~$x=-$,}	\\
							 &\bot	&&\text{otherwise,}
							\end{aligned}
						\right.
					\end{aligned}
					&\qquad&&
					\begin{aligned}
						\daTransFunc\tuple{\bot,(x,y)}&=\bot,
						\\
						\daTransFunc\tuple{\bot,\bot}&=\bot,
						\\
						\daTransFunc\tuple{\top,\top}=
						\daTransFunc\tuple{\emptyset,\top}&=\top,
						\\
						\daTransFunc\tuple{\top,\bot}=
						\daTransFunc\tuple{\emptyset,\bot}&=\bot,
					\end{aligned}
				\end{aligned}
			\]
			for~$x,x',y,y'\in\set{a,b,-}$,
			and for completeness,
			all images of~$\daTransFunc$ that are not specified above
			(because they correspond to useless transitions)
			are sent to state~$\bot$.
	\end{itemize}

  We make two further observations.
  First, notice that some states in~$\daStateSet$
  cannot be reached,
  namely the states~$(a,a)$ and~$(b,b)$,
  since receiving a letter implies not expecting it anymore.
  Thus, these two states can be eliminated from~$\daStateSet$
  without changing the accepted language.
  Second,~$\daAutomaton$ is not quasi-acyclic.
  Indeed,
  a node may for instance switch
  from state~$(-,-)$ to state~$(-,b)$ and then again to~$(-,-)$,
  as can be seen on the last node of the dipath~$aabbac$
  in Figure~\ref{fig:running-da-runs}.
  Nevertheless,
  as seen in Section~\ref{sec:cm-to-da},
  from the sumless counter machine
  of Example~\ref{ex:running-cm},
  we can construct
  an equivalent
  quasi-acyclic distributed automaton
  (see Figure~\ref{fig:cm2da}).
\end{uExample}

\paragraph*{Link with one-way cellular automata}

\begin{lemma}
  Let~$\daAutomaton$
  be a quasi-acyclic distributed automaton
  with~$k$ states
  and~$\Word=\Node_1\Node_2\cdots\Node_n$
  be a labeled \dipath\ of length~$n$.
  Then,
  the run~$\daRun=\tuple{\daRun_0,\daRun_1,\dots}$
  of~$\daAutomaton$ on~$\Word$
  satisfies:
  for each~$i\in\range[1]{n}$
  and each~$j>ki$
  we have~$\daRun_j(\Node_i)=\daRun_{j-1}(\Node_i)$.
\end{lemma}
\begin{proof}
  We proceed by induction on $i$.
  If~$i=1$,
  as the first node always receives the information~$\emptyset$,
  it evolves only dependently of its current state.
  Hence, after at most~$k+1$ steps
  some repetitions of state occurs.
  By quasi-acyclicity
  this repetition occurs at two successive time,
  while by determinism,
  the position stays in that repeated state forever.

  Let~$0<i<n$ be fixed,
  and suppose that for each~$j>ki$
  we have~$\daRun_j(\Node_i)=\daRun_{j-1}(\Node_i)$.
  We consider~the $(i+1)$-th node~$\Node_{i+1}$ of~$\Word$.
  By induction hypothesis,
  after~$ki$ initial steps,
  this node always receive the same information from its predecessor,
  namely, the state~$\daRun_{ki}(\Node_i)$.
  Hence,
  from that point,
  after at most~$k$ steps,
  the node~$\Node_{i+1}$ enters a state
  which is repeated at the next step,
  by quasi-acyclicity of~$\daAutomaton$.
  Finally,
  by determinism,
  this state is repeated forever.
  In other words,
  $\daRun_{j}(\Node_{i+1})=\daRun_{k(i+1)}(\Node_{i+1})$
  for each~$j>k(i+1)$.
  \qed
\end{proof}

\begin{proposition}
  Quasi-acyclic distributed automata
  are a special case of
  (one-dimensional, reversed)
  one-way cellular automata working in linear time.
\end{proposition}

\section{Translating between counter machines}
\label{app:cm-to-cm}

\begin{uProposition}[\ref{prp:copyless-to-sumless}]
  For every copyless $\cmCounterNum$-counter machine
  with $\cmAccess$-access,
  we can effectively construct an equivalent
  sumless $(2^\cmCounterNum)$-counter machine
  with $(\cmCounterNum \cdot \cmAccess)$-access.
\end{uProposition}
\begin{proof}
  \newcommand*{\vars}{\operatorname{vars}}
  \newcommand*{\const}{\operatorname{const}}
  The idea is simply to introduce a dedicated counter
  for each subset of counters $\cmCounterSubset$
  of the original machine~$\cmMachine$,
  and use this dedicated counter
  to store the sum of values of the counters in~$\cmCounterSubset$.
  Call this sum the \emph{value} of~$\cmCounterSubset$.
  Since $\cmMachine$ is copyless,
  it uses each of its counters at most once
  in any update function~$\cmUpdateFunc$.
  Therefore,
  the next value of $\cmCounterSubset$
  with respect to~$\cmUpdateFunc$
  can be expressed in terms of the current value of
  some other subset $\cmCounterSubset'$
  and a constant between
  $-\card{\cmCounterSubset} \cdot \cmAccess$
  and~$\card{\cmCounterSubset} \cdot \cmAccess$.
  This allows us to derive from~$\cmUpdateFunc$
  a sumless update function~$\cmUpdateFunc'$
  that operates on subsets of counters
  and uses constants in
  $\range[-\cmCounterNum \cmAccess]{\cmCounterNum \cmAccess}$.

  Formally,
  let
  $\cmMachine = \tuple{\cmStateSet,\cmCounterSet,\cmInitState,\cmTransFunc,\cmAcceptSet}$
  be a copyless machine with $\cmAccess$-access,
  over the alphabet~$\Alphabet$.
  We construct the sumless machine
  $\cmMachine' = \tuple{\cmStateSet,\cmCounterSet',\cmInitState,\cmTransFunc',\cmAcceptSet}$
  with $(\cmCounterNum \cmAccess)$-access
  such that
  $\cmCounterSet' = \powerset{\cmCounterSet}$
  and the result of the transition
  $\cmTransFunc'(\cmState[1], \cmTruncatedValuation', \Symbol)$
  is defined as follows,
  for
  any state
  $\cmState[1] \in \cmStateSet$,
  any $(\cmCounterNum \cmAccess)$-truncated valuation
  $\cmTruncatedValuation' \in
   \range[-\cmCounterNum\cmAccess]{\cmCounterNum\cmAccess}^{\cmCounterSet'}$\!,
  and any symbol
  $\Symbol \in \Alphabet$.
  To simplify the formalization,
  let $\cmTruncatedValuation$
  be the $\cmAccess$-truncated valuation of $\cmMachine$
  represented by~$\cmTruncatedValuation'$,
  that is,
  $\cmTruncatedValuation(\cmCounter) =
   \cut{-\cmAccess}{+\cmAccess}(\cmTruncatedValuation'(\set{\cmCounter}))$
  for $\cmCounter \in \cmCounterSet$,
  and assume that
  $\cmTransFunc(\cmState[1], \cmTruncatedValuation, \Symbol) =
   \tuple{\cmState[2], \cmUpdateFunc}$.
  Based on this,
  we define
  $\cmTransFunc'(\cmState[1], \cmTruncatedValuation', \Symbol) =
   \tuple{\cmState[2], \cmUpdateFunc'}$,
  where for
  $\cmCounterSubset \subseteq \cmCounterSet$,
  \begin{equation*}
    \cmUpdateFunc'(\cmCounterSubset) =
    \bigcup_{\cmCounter[1] \in \cmCounterSubset} \vars \bigl( \cmUpdateFunc(\cmCounter[1]) \bigr)
    \,+\,
    \sum_{\cmCounter[1] \in \cmCounterSubset} \const \bigl( \cmUpdateFunc(\cmCounter[1]) \bigr).
  \end{equation*}
  Here,
  $\vars \bigl( \cmUpdateFunc(\cmCounter[1]) \bigr)$
  denotes the set of counter variables
  that occur in the expression
  $\cmUpdateFunc(\cmCounter[1])$,
  and similarly,
  $\const \bigl( \cmUpdateFunc(\cmCounter[1]) \bigr)$
  denotes the constant
  $\cmConstant \in \range[-\cmAccess]{\cmAccess}$
  that occurs in that expression.
  \qed
\end{proof}

\begin{uProposition}[\ref{prp:nonnegative-counters}]
  For every sumless $\cmCounterNum$-counter machine
  with $\cmAccess$-access,
  we can effectively construct an equivalent machine
  that is also sumless with $\cmCounterNum$ counters and $\cmAccess$-access,
  but whose counters never store any negative values
  (regardless of the input word).
\end{uProposition}
\begin{proof}
  \newcommand*{\cmSignFunc}[1][]{\ifthenelse{\isempty{#1}\OR\equal{#1}{1}}{\alpha}{\beta}}
  It suffices to represent
  each counter~$\cmCounter$ of the original machine
  in such a way that
  the absolute value of~$\cmCounter$ is stored in a counter
  and its sign is remembered in finite-state memory.
  Since the machine is sumless,
  we do not have to deal with
  the problem of computing
  the sum of a positive and a negative counter value.

  For the sake of completeness,
  let us perform a formal construction.
  Given the sumless machine
  $\cmMachine = \tuple{\cmStateSet,\cmCounterSet,\cmInitState,\cmTransFunc,\cmAcceptSet}$
  with $\cmAccess$-access
  over the alphabet~$\Alphabet$,
  we construct the sumless machine
  $\cmMachine' = \tuple{\cmStateSet',\cmCounterSet,\cmInitState',\cmTransFunc',\cmAcceptSet'}$
  such that
  $\cmStateSet' = \cmStateSet \times \set{-1,1}^\cmCounterSet$,
  $\cmInitState' = \bigtuple{\cmInitState, \setbuilder{\cmCounter \mapsto 1}{\cmCounter \in \cmCounterSet}}$,
  $\cmAcceptSet' = \cmAcceptSet \times \set{-1,1}^\cmCounterSet$,
  and the outcome of the transition
  $\cmTransFunc'\bigl( \tuple{\cmState[1], \cmSignFunc[1]}, \cmTruncatedValuation', \Symbol \bigr)$
  is defined as follows,
  for
  any state
  $\tuple{\cmState[1], \cmSignFunc[1]} \in \cmStateSet'$,
  any \mbox{$\cmAccess$-truncated} valuation
  $\cmTruncatedValuation' \in
   \range[0]{\cmAccess}^\cmCounterSet$\!,
  and any symbol
  $\Symbol \in \Alphabet$.
  Assume that
  $\cmTransFunc(\cmState[1], \cmTruncatedValuation, \Symbol) =
   \tuple{\cmState[2], \cmUpdateFunc}$,
  where
  $\cmTruncatedValuation \in
   \range[-\cmAccess]{\cmAccess}^\cmCounterSet$
  is the $\cmAccess$-truncated valuation of $\cmMachine$
  represented by~$\cmSignFunc[1]$ and~$\cmTruncatedValuation'$,
  \ie,
  $\cmTruncatedValuation(\cmCounter) =
   \cmSignFunc[1](\cmCounter) \cdot \cmTruncatedValuation'(\cmCounter)$
  for $\cmCounter \in \cmCounterSet$.
  Based on this,
  we define
  $\cmTransFunc'\bigl( \tuple{\cmState[1], \cmSignFunc[1]}, \cmTruncatedValuation', \Symbol \bigr) =
   \bigtuple{\tuple{\cmState[2], \cmSignFunc[2]}, \cmUpdateFunc'}$
  such that
  for every $\cmCounter[1] \in \cmCounterSet$
  with $\cmUpdateFunc(\cmCounter[1]) = \cmConstant$,
  \begin{alignat*}{4}
      \cmSignFunc[2](\cmCounter[1]) &= 1
    &\quad 
      &\text{and}
    &\quad 
      \cmUpdateFunc'(\cmCounter[1]) &= \cmConstant
    &\qquad 
      &\text{if $\cmConstant \geq 0$,}
    \\[1ex] 
      \cmSignFunc[2](\cmCounter[1]) &= -1
    & 
      &\text{and}
    & 
      \cmUpdateFunc'(\cmCounter[1]) &= -\cmConstant
    & 
      &\text{otherwise,}
  \end{alignat*}
  and
  for every $\cmCounter[1] \in \cmCounterSet$
  with $\cmUpdateFunc(\cmCounter[1]) = \cmCounter[2] + \cmConstant$,
  \begin{alignat*}{4}
      \cmSignFunc[2](\cmCounter[1]) &= \cmSignFunc[1](\cmCounter[2])
    &\quad 
      &\text{and}
    &\quad 
      \cmUpdateFunc'(\cmCounter[1]) &=
      \cmCounter[2] + \cmSignFunc(\cmCounter[2]) \cdot \cmConstant
    &\qquad 
      &\text{%
         if $\cmTruncatedValuation'(\cmCounter[2])
             + \cmSignFunc[1](\cmCounter[2]) \cdot \cmConstant \geq 0$,
       }
    \\[1ex] 
      \cmSignFunc[2](\cmCounter[1]) &= -\cmSignFunc[1](\cmCounter[2])
    & 
      &\text{and}
    & 
      \cmUpdateFunc'(\cmCounter[1]) &=
      \abs{\cmConstant} - \cmTruncatedValuation'(\cmCounter[2])
    & 
      &\text{otherwise.}
  \end{alignat*}
  There are no other cases to consider,
  since $\cmMachine$ does not compute sums of multiple counters.
  \qed
\end{proof}

\begin{uProposition}[\ref{prp:1-access}]
  For every $\cmCounterNum$-counter machine $\cmMachine$ with $\cmAccess$-access,
  we can effectively construct an equivalent
  $(\cmAccess \cdot \cmCounterNum)$-counter machine $\cmMachine'$ with $1$-access.
  If $\cmMachine$ is copyless or sumless,
  then so is $\cmMachine'$.
  Furthermore,
  if $\cmMachine$ is sumless,
  then $\cmMachine'$ requires only $\cmCounterNum$ counters.
\end{uProposition}
\begin{proof}
  \newcommand*{\ModEnc}[2]{\langle#1\rangle_{#2}}
  \newcommand*{\ModEncTuple}[2]{\tuple{#1:#2}}
  \newcommand*{\BigModEncTuple}[2]{\bigtuple{#1:#2}}
  \newcommand*{\Remainder}[1]{r_{#1}}
  \newcommand*{\Plus}{\varoplus}
  \newcommand*{\cmRemainderFunc}[1][]{\ifthenelse{\isempty{#1}\OR\equal{#1}{1}}{\alpha}{\beta}}
  \newcommand*{\cmTranslation}[1]{T_{#1}}
  The key idea is that $\cmMachine'$ represents
  each counter $\cmCounter$ of $\cmMachine$
  by $\cmAccess$ counters
  $\cmCounter_0, \dots, \cmCounter_{\cmAccess-1}$
  over which the value of $\cmCounter$
  is distributed as uniformly as possible.
  That is,
  the value of $\cmCounter$
  is equal to the sum of the values of
  $\cmCounter_0, \dots, \cmCounter_{\cmAccess-1}$,
  and any two of the latter values differ by at most $1$.
  To this end,
  we first make the following observations.

  We can represent any integer $n \in \Integer$
  as an $(\cmAccess + 1)$-tuple
  $\ModEnc{n}{\cmAccess} =
   {\ModEncTuple{n_0, \dots, n_{\cmAccess-1}}{\Remainder{n}}}$,
   where
  \begin{equation*}
    n_i \defeq
    \begin{cases*}
      \ceil{n/\cmAccess}  & if $i < \Remainder{n}$ \\
      \floor{n/\cmAccess} & otherwise
    \end{cases*}
    \quad\text{and}\qquad
    \Remainder{n} \defeq n \bmod \cmAccess,
  \end{equation*}
  for $0 \leq i < \cmAccess$.
  Note that $\sum_{0 \leq i < \cmAccess}(n_i) = n$.
  We now define an addition operator~$\Plus$
  on such tuple representations
  that is consistent with the usual addition on integers.
  Consider
  $\ell,m \in \Integer$
  such that
  $\ell + m = n$,
  and let
  $\ModEnc{\ell}{\cmAccess} =
   \ModEncTuple{\ell_0, \dots, \ell_{\cmAccess-1}}{\Remainder{\ell}}$
  and
  $\ModEnc{m}{\cmAccess} =
   \ModEncTuple{m_0, \dots, m_{\cmAccess-1}}{\Remainder{m}}$.
  We require that
  $\ModEnc{\ell}{\cmAccess} \Plus \ModEnc{m}{\cmAccess} \defeq
   \ModEnc{\ell+m}{\cmAccess} =
   \ModEnc{n}{\cmAccess}$.
  Rather conveniently,
  $\Plus$ can be evaluated directly on the tuples
  $\ModEnc{\ell}{\cmAccess}$ and $\ModEnc{m}{\cmAccess}$,
  without first computing the represented integers $\ell$~and~$m$:
  it is a routine exercise to verify that
  \begin{equation}
    n_i = \ell_i + m_{((i-\Remainder{\ell})\bmod\cmAccess)}
    \qquad\text{and}\qquad
    \Remainder{n} = (\Remainder{\ell} + \Remainder{m}) \bmod \cmAccess.
    \tag{$\ast$} \label{eq:tuple-addition}
  \end{equation}
  The significant point here is
  that it suffices to know the value of $\Remainder{\ell}$
  (or, by symmetry, $\Remainder{m}$)
  in order to determine
  which sums of the form $\ell_i + m_j$
  yield the components of $\ModEnc{n}{\cmAccess}$.
  We do not need to know the values of $\ell_i$ and $m_j$,
  only be able to compute their sum.
  Hence,
  it makes sense to extend the domain of definition of $\Plus$
  to tuples of the form
  $\ModEncTuple{\cmExpression_0, \dots, \cmExpression_{\cmAccess-1}}{\Remainder{\cmExpression}}$,
  where
  $\cmExpression_0, \dots, \cmExpression_{\cmAccess-1}$
  are counter expressions
  and $\Remainder{\cmExpression}$ is an integer in
  $\range[0]{\cmAccess - 1}$.
  On such tuples,
  the definition of $\Plus$
  is completely analogous to \eqref{eq:tuple-addition},
  we simply use counter expressions instead of integer values.
  (It does not matter that this extended version of $\Plus$ is not commutative.)

  Coming back to our actual goal,
  let $\cmCounterSet$ be the set of counter variables of $\cmMachine$
  and assume we are given a function
  $\cmRemainderFunc \colon \cmCounterSet \to \range[0]{\cmAccess - 1}$.
  The idea is that
  $\cmRemainderFunc$ will be stored in the finite-state memory of~$\cmMachine'$
  and satisfy
  $\cmRemainderFunc(\cmCounter) = (\cmValuation(\cmCounter) \bmod \cmAccess)$
  for the current valuation~$\cmValuation$ of~$\cmCounterSet$.
  For each counter $\cmCounter \in \cmCounterSet$,
  the machine $\cmMachine'$ will have $\cmAccess$ counters
  $\cmCounter_0, \dots, \cmCounter_{\cmAccess-1}$
  valuated by $\cmValuation'$
  such that
  $\ModEnc{\cmValuation(\cmCounter)}{\cmAccess} =
   \BigModEncTuple{\cmValuation'(\cmCounter_0), \dots, \cmValuation'(\cmCounter_{\cmAccess-1})}
                  {\cmRemainderFunc(\cmCounter)}$.
  To implement this,
  we define the following function $\cmTranslation{\cmRemainderFunc}$,
  which translates
  each counter expression~$\cmExpression$ of~$\cmMachine$
  to the corresponding counter expressions
  $\cmExpression_0, \dots, \cmExpression_{\cmAccess-1}$
  used by $\cmMachine'$ to simulate $\cmExpression$
  with respect to $\cmRemainderFunc$:
  \begin{equation*}
    \cmTranslation{\cmRemainderFunc}(\cmCounter + \cmExpression) =
    \BigModEncTuple{\cmCounter_0, \dots, \cmCounter_{\cmAccess-1}}{\cmRemainderFunc(\cmCounter)}
    \Plus\:\! \cmTranslation{\cmRemainderFunc}(\cmExpression)
    \qquad\text{and}\qquad
    \cmTranslation{\cmRemainderFunc}(\cmConstant) =
    \ModEnc{\cmConstant}{\cmAccess},
  \end{equation*}
  for $\cmCounter \in \cmCounterSet$
  and $\cmConstant \in \range[-\cmAccess]{\cmAccess}$.
  Notice that a counter expression containing
  a constant between $-\cmAccess$ and $\cmAccess$
  gets translated to $\cmAccess$ expressions
  with constants between $-1$ and $1$.

  Let us now formally construct the machine $\cmMachine'$.
  Given
  $\cmMachine = \tuple{\cmStateSet,\cmCounterSet,\cmInitState,\cmTransFunc,\cmAcceptSet}$
  with $\cmAccess$-access
  over the alphabet~$\Alphabet$,
  we define
  $\cmMachine' = \tuple{\cmStateSet',\cmCounterSet',\cmInitState',\cmTransFunc',\cmAcceptSet'}$
  with $1$-access
  such that
  \begin{alignat*}{2}
    \cmStateSet'   &= \cmStateSet \times \range[0]{\cmAccess - 1}^\cmCounterSet, &\qquad
    \cmCounterSet' &= \setbuilder{\cmCounter_i}{\text{$\cmCounter \in \cmCounterSet$ and $0 \leq i < \cmAccess$}}, \\
    \cmInitState'  &= \bigtuple{\cmInitState, \setbuilder{\cmCounter \mapsto 0}{\cmCounter \in \cmCounterSet}}, &
    \cmAcceptSet'  &= \cmAcceptSet \times \range[0]{\cmAccess - 1}^\cmCounterSet,
  \end{alignat*}
  and the result of
  $\cmTransFunc'\bigl( \tuple{\cmState[1], \cmRemainderFunc[1]}, \cmTruncatedValuation', \Symbol \bigr)$
  is defined as follows,
  for
  any state
  $\tuple{\cmState[1], \cmRemainderFunc[1]} \in \cmStateSet'$,
  any $1$-truncated valuation
  $\cmTruncatedValuation' \in \range[-1]{1}^{\cmCounterSet'}$\!,
  and any symbol
  $\Symbol \in \Alphabet$.
  Let $\cmTruncatedValuation \in \range[-\cmAccess]{\cmAccess}^\cmCounterSet$
  be the corresponding \mbox{$\cmAccess$-truncated} valuation of $\cmMachine$
  that is encoded
  by~$\cmRemainderFunc[1]$ and~$\cmTruncatedValuation'$, \ie,
  \begin{equation*}
    \cmTruncatedValuation(\cmCounter) =
    \begin{cases*}
      \cmAccess
        & if $\cmTruncatedValuation'(\cmCounter_{\cmAccess-1}) = 1$, \\
      \cmRemainderFunc[1](\cmCounter)
        & if $\cmTruncatedValuation'(\cmCounter_{\cmAccess-1}) = 0$ \\
      \cmRemainderFunc[1](\cmCounter) - \cmAccess
        & if $\cmTruncatedValuation'(\cmCounter_{\cmAccess-1}) = -1$
          and $\cmTruncatedValuation'(\cmCounter_0) = 0$, \\
      -\cmAccess
        & otherwise,
    \end{cases*}
    \tag{$\dagger$} \label{eq:truncated-valuation}
  \end{equation*}
  for $\cmCounter \in \cmCounterSet$,
  and assume that
  $\cmTransFunc(\cmState[1], \cmTruncatedValuation, \Symbol) =
   \tuple{\cmState[2], \cmUpdateFunc}$.
  Then we have
  $\cmTransFunc'\bigl( \tuple{\cmState[1], \cmRemainderFunc[1]}, \cmTruncatedValuation', \Symbol \bigr) =
   \bigtuple{\tuple{\cmState[2], \cmRemainderFunc[2]}, \cmUpdateFunc'}$,
  where for
  $0 \leq i < \cmAccess$
  and
  $\cmCounter \in \cmCounterSet$
  with
  $\cmTranslation{\cmRemainderFunc[1]}\bigl( \cmUpdateFunc(\cmCounter) \bigr) =
   \ModEncTuple{\cmExpression_0, \dots, \cmExpression_{\cmAccess-1}}{\Remainder{\cmExpression}}$,
  \begin{equation*}
    \cmRemainderFunc[2](\cmCounter) = \Remainder{\cmExpression}
    \qquad\text{and}\qquad
    \cmUpdateFunc'(\cmCounter_i) = \cmExpression_i.
  \end{equation*}

  Now it is straightforward to see that
  if $\cmMachine$ is copyless or sumless,
  then so is $\cmMachine'$.
  In particular,
  if $\cmMachine$ is sumless,
  then for all $\cmCounter[1] \in \cmCounterSet$,
  the expression $\cmUpdateFunc(\cmCounter[1])$
  is either $\cmConstant$~or~${\cmCounter[2] + \cmConstant}$,
  for some
  $\cmConstant \in \range[-\cmAccess]{\cmAccess}$
  and $\cmCounter[2] \in \cmCounterSet$.
  By looking at our construction of~$\cmMachine'$,
  we can observe that
  this implies that for each $i \in \range[0]{\cmAccess-1}$,
  the expression
  $\cmUpdateFunc'(\cmCounter[1]_i)$
  is either $\cmConstant'$ or $\cmCounter[2]_i + \cmConstant'$,
  for some $\cmConstant' \in \range[-1]{1}$.
  This means that the counters in
  $\setbuilder{\cmCounter_i}{\cmCounter \in \cmCounterSet}$
  are updated independently of the counters in
  $\setbuilder{\cmCounter_j}{\cmCounter \in \cmCounterSet,\, j \neq i}$.
  Furthermore,
  we may assume by Proposition~\ref{prp:nonnegative-counters}
  that a sumless machine
  never stores any negative values in its counters,
  and therefore
  only the first two cases of equation~\eqref{eq:truncated-valuation}
  are relevant.
  As can be seen there,
  although $\cmMachine'$ writes to all of its counters,
  it reads only the last counter $\cmCounter[1]_{\cmAccess-1}$
  for each $\cmCounter[1] \in \cmCounterSet$.
  Hence,
  if~$\cmMachine$ is sumless,
  the remaining counters
  $\cmCounter[1]_0, \dots, \cmCounter[1]_{\cmAccess-2}$
  are completely useless and can thus be~removed.

  As a final remark,
  notice that the values of
  the $\cmAccess$ counters representing
  $\cmCounter \in \cmCounterSet$
  will never differ by more than $1$.
  The largest value is stored in $\cmCounter_0$
  and the smallest (possibly equal to the largest)
  in $\cmCounter_{\cmAccess-1}$.
  Since the remaining counters
  $\cmCounter_1, \dots, \cmCounter_{\cmAccess-2}$
  always contain one of those two values,
  they are in principle redundant.
  That is,
  we can easily convert $\cmMachine'$ into an equivalent
  $2\cmCounterNum$-counter machine $\cmMachine''$ with $1$-access.
  The reason we do not use this optimized construction
  is that it does not preserve copylessness.
  \qed
\end{proof}
\section{From counter machines to distributed automata}
\label{app:cm-to-da}

\begin{uProposition}[\ref{prp:cm-to-da}]
  For every sumless $\cmCounterNum$-counter machine with $1$-access,
  we can effectively construct
  an equivalent quasi-acyclic distributed automaton
  with at most $(\cmCounterNum+2)$ loops per~trace.
\end{uProposition}
\begin{proof}
  \providecommand*\subsubparagraph[1]{\emph{#1}}%
  \newcommand*\cmtodaDelay{\ensuremath\top}%
  \newcommand*\cmtodaCount{\ensuremath1}%
  \newcommand*\cmtodaEOCount{\ensuremath0}%
  \newcommand*\cmtodaEOCountbis{\ensuremath\bot}%
  \newcommand*\cmtodaQuiescent{\ensuremath\sharp}%
  \newcommand*\cmtodaVector{\ensuremath y}%
  Let
  $\cmMachine=\tuple{\cmStateSet,\cmCounterSet,\cmInitState,\cmTransFunc,\cmAcceptSet}$
  be a sumless $\cmCounterNum$-counter machine
  with $1$-access
  over the alphabet~$\Alphabet$.
  By Proposition~\ref{prp:nonnegative-counters},
  we may assume without loss of generality
  that $\cmMachine$ never stores any negative values in its counters.
  We build a quasi-acyclic distributed automaton
  $\daAutomaton = \tuple{\daStateSet,\daInitFunc,\daTransFunc,\daAcceptSet}$
  with at most $(\cmCounterNum+2)$ loops per trace,
  which simulates $\cmMachine$ on every nonempty input
  viewed as a labeled \dipath{}.
  Our simulation is based on an exchange
  of time and space in the following sense:
  given a run $\cmRun$ of $\cmMachine$
  on some input word $\Word\in\Alphabet^+$,
  we simulate it by a run $\daRun$ of $\daAutomaton$
  over the labeled \dipath~$\Word$,
  in which the \defd{history} of the $i$-th node,
  \ie, the sequence of states entered at position~$i$, (space)
  is an encoding of the $(i+1)$-th memory configuration (time) in $\cmRun$,
  for each position $i$
  (%
  the initial configuration of $\cmMachine$ is not encoded%
  )%
  .
  \smallbreak

  We actually encode more than a sequence of memory configurations:
  at each position, the history additionally carries
  the input letter labeling the position
  and the register update function
  that have been applied to enter the encoded configuration.
  We refer to these augmented configuration
  as \defd{history memory configuration}.
  Moreover,
  a delay is prepended to each history,
  in order to leave time to information to transit
  from the leftmost position towards the rightmost one.
  As a side effect, this delay yields non-unicity
  of the encoding of a history memory configuration:
  two different histories may encode the same history memory configuration
  as the delay depends on the position of the input node.
  More precisely,
  some states of~\daAutomaton
  are identified as \emph{delaying states},
  and at position~$i$, $1\leq i\leq n$,
  the history starts with~$(2i-1)$ repetitions
  of one of these delaying\nolinebreak\xspace states.
  \smallbreak

  Our encoding of history memory configurations
  uses $(\cmCounterNum+1)$ \emph{tracks},
  that is,
  the states of $\daAutomaton$
  are tuples from the direct product of $(\cmCounterNum+1)$ finite sets
  and a \emph{track} is defined as the projection
  over one specific component
  of the history at some position.
  Each $i$-th track of a history of $\daAutomaton$ for $1\leq i\leq k$
  (namely, the \emph{counter tracks})
  encodes the value of the $i$-th counter of $\cmMachine$,
  while the last track
  (namely, the \emph{transition track})
  initially contains the label of the current position,
  and then contains the state of the configuration,
  together with the counter update function
  that was lastly performed.

  \paragraph{Counter tracks.}
  Intuitively,
  a counter track contains the corresponding counter value
  written in unary using the symbol $\cmtodaCount$.
  For technical reasons, some markers should be added:
  first, we allow to delay the encoding
  by prepending a positive number of symbol $\cmtodaDelay$;
  second we close the encoding by appending two endmarkers
  $\cmtodaEOCount$ and $\cmtodaEOCountbis$,
  both of them occurring only once in each counter track;
  lastly, we append an infinite repetition of the symbol $\cmtodaQuiescent$.
  Hence, the counter track corresponding to a counter name $\cmCounter\in\cmCounterSet$
  under valuation $\cmValuation$
  is of the form:
  \begin{equation}
    \label{eq:counter track}
    \cmtodaDelay^{d+1}\cdot
    \cmtodaCount^{\cmValuation(\cmCounter)}\cdot
    \cmtodaEOCount\cdot
    \cmtodaEOCountbis\cdot
    \cmtodaQuiescent^\omega
  \end{equation}
  where
  $d$ is a positive integer
  which is shared by every counter track of a history,
  and to which we refer as \emph{the delay} of the history.

  \paragraph{Transition track.}
  The transition track crosses two phases.
  First,
  it simply contains the letter that labels the current position.
  Then,
  it switches to the result $\tuple{\cmState,\cmUpdateFunc}$
  of the simulated transition.
  Intuitively,
  this tracks stores the input letter
  until it can determine which transition to perform
  (when the required information has been transmitted by the preceding position)
  and then enters the result of this transition.
  Therefore, at any position, the transition track has the following form:
  \begin{equation}
    \label{eq:transition track}
    \Symbol^d\cdot
    \tuple{\cmState,\cmUpdateFunc}^\omega
  \end{equation}
  for some state $\cmState$ of $\cmMachine$
  and some counter update function $\cmUpdateFunc$.
  Here, $d$ is the delay of the history,
  \ie, the length of the maximal prefix in $\cmtodaDelay^*$
  of the counter tracks of the history minus $1$.

  \paragraph{Next history.}
  Suppose now that the history at a given position $i$
  has the form
  given by track
  in~\eqref{eq:counter track}
  and~\eqref{eq:transition track}.
  If the position $(i+1)$ exists
  and is labeled by $\Symbol'\in\Alphabet$
  and if
  $\cmTransFunc(\cmState,\,\cut{0}{1}\circ\cmValuation,\,\Symbol')=\tuple{\cmState',\cmUpdateFunc'}$,
  the history at position $(i+1)$
  will have counter track associated to a counter name $\cmCounter$
  \begin{equation*}
    \cmtodaDelay^{d+3}\cdot
    \cmtodaCount^{\cmExtendedValuation\circ\cmUpdateFunc'(\cmCounter)}\cdot
    \cmtodaEOCount\cdot
    \cmtodaEOCountbis\cdot
    \cmtodaQuiescent^\omega
  \end{equation*}
  where $\cmExtendedValuation$
  is the extended valuation obtained from the valuation $\cmValuation$,
  and transition track
  \begin{equation*}
    \Symbol'^{d+2}\cdot
    \tuple{{\cmUpdateFunc}',\cmState'}^\omega
  \end{equation*}

  \subsubparagraph{Remark about delay.}
  Though it is clear
  that increasing the delay by $1$ at each position
  is required
  in order to leave enough time
  for the leftmost position to influence the rightmost position
  (which is the pointed position)
  by transferring at least a state,
  we actually increase it by $2$ at each position,
  in order to be able to decrement counters,
  as we will describe later.
  \smallbreak

  The next paragraphs are devoted
  to the definition of
  the initialization function $\daInitFunc$ and
  the transition function $\daTransFunc$
  of $\daAutomaton$,
  in order to obtain the successive history described above.
  For clarity,
  the history are described
  by splitting them into several phases.

  \paragraph{Initializing the states.}
  The state of each position
  is initialized as the tuple
  $\tuple{\cmtodaDelay^k,\Symbol}$
  where~$\Symbol$ is the letter labeling the position
  (remember that~$k$ is the number of counters of~\cmMachine).
  Formally:
  \begin{equation*}
    \daInitFunc(\Symbol)=\tuple{\cmtodaDelay^k,\Symbol}
    \qquad
    \text{for each $\Symbol\in\Alphabet$}
  \end{equation*}

  \paragraph{Delaying.}
  After initialization,
  the position increase the delay
  with respect to the previous position,
  whenever it exists.
  That is:
  \begin{align*}
    &\daTransFunc\big((\cmtodaDelay^k,\Symbol),\, (\cmtodaDelay^k,\Symbol')\big)
    =
    (\cmtodaDelay^k,\Symbol')
    &&
    \text{for each $\Symbol,\Symbol'\in\Alphabet$}
    \\
    &\daTransFunc\big((\cmtodaDelay^k,\tuple{\cmState,\cmUpdateFunc}),\, (\cmtodaDelay^k,\Symbol')\big)
    =
    (\cmtodaDelay^k,\Symbol')
    &&
    \text{%
      for each $\Symbol'\in\Alphabet$,
      $\cmState\in\cmStateSet$
      and
      $\cmUpdateFunc\in\cmExpressionSet(\cmCounterSet,1)^\cmCounterSet$
    }
  \end{align*}

  \paragraph{Determining the transition.}
  After the delaying phase,
  assuming a preceding node exists,
  the current node receives,
  for the first time,
  a state of the form $\tuple{\cmtodaVector,\tuple{\cmState,\cmUpdateFunc}}$
  for some vector $\cmtodaVector$ of $\set{\cmtodaCount,\cmtodaEOCount}^k$.
  Observe that
  the $0$'s of $y$ correspond exactly
  to the counter track components
  that are associated to the counters
  which have value $0$ in the history memory configuration
  encoded at the preceding position.
  Therefore,
  and since the label $\Symbol$
  of the current position is still available
  as being part of the current state,
  $\daAutomaton$
  can compute
  $\cmTransFunc(\cmState,\cmtodaVector,\Symbol)=\tuple{\cmState',\cmUpdateFunc'}$.
  Similarly,
  if the node is the first one
  (and hence receives constantly $\daNoState$),
  $\daAutomaton$
  can compute
  $\cmTransFunc(\cmInitState,\cmInitValuation,\Symbol)=\tuple{\cmState',\cmUpdateFunc'}$.
  In both cases,
  $\daAutomaton$
  enters the state
  $\tuple{\cmtodaDelay^k,\tuple{\cmState',\cmUpdateFunc'}}$.
  Formally:
  \begin{align*}
    \daTransFunc\big(
    \tuple{\cmtodaVector,\tuple{\cmState,\cmUpdateFunc}},\,
    \tuple{\cmtodaDelay^k,\Symbol}
    \big)
    &=
    \tuple{\cmtodaDelay^k,\tuple{{\cmUpdateFunc}',\cmState'}}
    \\&
    \begin{gathered}
      \text{%
        for each $y\in\set{0,1}^k$,
        $\cmUpdateFunc,{\cmUpdateFunc}'\in\cmExpressionSet(\cmCounterSet,1)^\cmCounterSet$,
        $\cmState,\cmState'\in\cmStateSet$%
      }\\
      \text{%
        and $\Symbol\in\Alphabet$
      }
      \text{%
        such that
        $\cmTransFunc(\cmState,\cmtodaVector,\Symbol)=\tuple{{\cmUpdateFunc}',\cmState'}$%
      }
    \end{gathered}
    \\[8pt]
    \daTransFunc\big(
    \daNoState,\,
    \tuple{\cmtodaDelay^k,\Symbol}
    \big)
    &=
    \tuple{\cmtodaDelay^k,\tuple{\cmState',\cmUpdateFunc'}}
    \\&
    \begin{gathered}
      \text{%
        for each $a\in\Alphabet$,
        $\cmUpdateFunc'\in\cmExpressionSet(\cmCounterSet,1)^\cmCounterSet$
        and $\cmState'\in\cmStateSet$%
      }
      \\
      \text{%
        such that
        $\cmTransFunc(\cmInitState,\cmInitValuation,\Symbol)=\tuple{\cmState',\cmUpdateFunc'}$%
      }
    \end{gathered}
  \end{align*}

  \paragraph{Updating the counters.}
  The last phase consists in updating the counters.
  This is done after having determined the transition as described previously.
  Therefore, the transition track component
  of the current state
  (and of every state entered at the current position after that time)
  stores, in particular,
  the update function $\cmUpdateFunc'$ to be performed.
  Since $\cmMachine$ is sumless,
  for every $\cmCounter[1]\in\cmCounterSet$
  either:
  (1) $\cmUpdateFunc'(\cmCounter[1])=\cmConstant$;
  or (2) $\cmUpdateFunc'(\cmCounter[1])=\cmCounter[2]+\cmConstant$,
  for some counter name $\cmCounter[2]\in\cmCounterSet$
  and some constant $\cmConstant$.
  Moreover,
  we suppose without loss of generality
  that a transition performed from the initial configuration
  has only counter update of the form (1).

  Let $\daState$ (\resp\ $\daState'$)
  be either $\daNoState$ if the current position is the first one
  or the state transmitted by the preceding position otherwise
  (%
  \resp\ the state at the current position,
  that includes $\cmUpdateFunc'$
  in its transition track component%
  )
  at some time of this last phase.
  Our goal is to define $\daTransFunc(\daState,\daState')$.
  For a counter $\cmCounter[1]$,
  we refer to the component of a state of $\daAutomaton$
  that corresponds to the counter track associated with $\cmCounter[1]$
  as its \emph{$\cmCounter[1]$-component}.
  We observe that the $\cmCounter[1]$-component of $\daTransFunc(\daState,\daState')$,
  denoted $\daTransFunc_{\cmCounter[1]}(\daState,\daState')$,
  depends only on $\cmUpdateFunc'$,
  on the content of the $\cmCounter[1]$-component of $\daState'$,
  denoted ${\daState}'_{\cmCounter[1]}$,
  and, possibly,
  on the $\cmCounter[2]$-component of $\daState$,
  denoted ${\daState}_{\cmCounter[2]}$,
  in the case
  $\daState\neq\daNoState$ and
  $\cmUpdateFunc'(\cmCounter[1])=\cmCounter[2]+\cmConstant$
  for some counter name $\cmCounter[2]$
  and some constant $\cmConstant\in\range[-1]{1}$,
  \ie, the form (2) described previously.
  In particular,
  counter tracks at a given position
  are pairwise independent,
  as far as the delay is fixed.
  This allows us to define the transition function of $\daAutomaton$
  by considering each counter track independently
  (%
  remember that
  the transition track component
  is kept unchanged forever
  from the end of the previous phase%
  ).

  We consider the track corresponding to the counter $\cmCounter[1]$.
  Let first assume
  that~$\daState'_{\cmCounter[1]}\in\set{\cmtodaDelay,\cmtodaCount}$.
  We proceed by case:
  \begin{enumerate}
  \item
    Suppose that $\cmUpdateFunc'(\cmCounter[1])=\cmConstant$
    for some $\cmConstant\in\range[-1]{1}$.
    In that case, we define:
    \begin{align*}
      \daTransFunc_{\cmCounter[1]}(\daState,\daState')
      &=
      \left\{
        \begin{array}{lcl}
          \cmtodaCount
          &&
          \text{if $\cmConstant=1$ and ${\daState}'_{\cmCounter[1]}=\cmtodaDelay$}
          \\
          \cmtodaEOCount
          &&
          \text{otherwise}
        \end{array}
      \right.
      &
      \fbox{%
        \text{%
          ${\daState}'_{\cmCounter[1]}\in\set{\cmtodaDelay,\cmtodaCount}$
          and
          ${\cmUpdateFunc}'(\cmCounter[1])=\cmConstant$
        }%
      }
    \end{align*}
  \item
    Consider now
    the case $\cmUpdateFunc'(\cmCounter[1])=\cmCounter[2]+\cmConstant$
    for some counter name $\cmCounter[2]$
    and some constant $\cmConstant\in\range[-1]{1}$.
    By assumption,
    this implies that $\daState\neq\daNoState$
    and therefore that $\daState_{\cmCounter[2]}$ is defined.
    In this case, we define:
    \begin{align*}
      \daTransFunc_{\cmCounter[1]}(\daState,\daState')
      &=
      \left\{
        \begin{array}{lcl}
          \cmtodaCount
          &&
          \text{if ${\daState}_{\cmCounter[2]}=\cmtodaCount$}
          \\
          \cmtodaCount
          &&
          \text{if ${\daState}_{\cmCounter[2]}=\cmtodaEOCount$ and $\cmConstant\geq0$}
          \\
          \cmtodaCount
          &&
          \text{if ${\daState}_{\cmCounter[2]}=\cmtodaEOCountbis$ and $\cmConstant=1$}
          \\
          \cmtodaEOCount
          &&
          \text{otherwise}
        \end{array}
      \right.
      &
      \fbox{%
        \text{%
          ${\daState}'_{\cmCounter[1]}\in\set{\cmtodaDelay,\cmtodaCount}$
          and
          ${\cmUpdateFunc}'(\cmCounter[1])=\cmCounter[2]+\cmConstant$
        }%
      }
    \end{align*}
  \end{enumerate}

  Finally,
  in the case~$\daState'_{\cmCounter[1]}\in\set{\cmtodaEOCount,\cmtodaEOCountbis,\cmtodaQuiescent}$,
  whatever the update to simulate is,
  when $\daState'_{\cmCounter[1]}$ is equal to $\cmtodaEOCount$
  (\resp\ $\cmtodaEOCountbis$ or $\cmtodaQuiescent$),
  it becomes $\cmtodaEOCountbis$ (\resp\ $\cmtodaQuiescent$)
  at the next step
  :
  \begin{align*}
    \daTransFunc_{\cmCounter[1]}(\daState,\daState')
    &=
    \left\{
      \begin{array}{lcl}
        \cmtodaEOCountbis
        &&
        \text{%
          if ${\daState}'_{\cmCounter[1]}=\cmtodaEOCount$%
        }
        \\
        \cmtodaQuiescent
        &&
        \text{%
          if ${\daState}'_{\cmCounter[1]}=\cmtodaEOCountbis$
          or ${\daState}'_{\cmCounter[1]}=\cmtodaQuiescent$%
        }
      \end{array}
    \right.
    &
    \fbox{%
      \text{%
        $\daState'_{\cmCounter[1]}\in\set{\cmtodaEOCount,\cmtodaEOCountbis,\cmtodaQuiescent}$%
      }
    }
  \end{align*}

  \paragraph{Accepting.}
  The accepting states of $\daAutomaton$ are defined
  as those tuples
  which contain an accepting state of $\cmMachine$
  in their transition track component.
  In that way,
  an nonempty input word is recognized by $\daAutomaton$
  if and only if
  it is accepted by $\cmMachine$,
  by construction.

  \paragraph{Number of loops per path.}
  We now evaluate the maximal number of looping state
  along a trace of $\daAutomaton$.
  During the delaying phase,
  every history enters one looping state
  (which is actually the initial state of the history).
  The second phase,
  consists in a single step,
  which therefore visit no looping state.
  Lastly, the counter updating phase
  can view each counter track independently switch
  from $\cmtodaCount$ to $\cmtodaQuiescent$
  (%
  through two sequential steps,
  namely the two steps
  that use the symbols $\cmtodaEOCount$
  and $\cmtodaEOCountbis$ respectively%
  )%
  ,
  while the transition track component remains constant.
  This yields at most $(k+1)$ looping states along a path
  during that phase
  (%
  the last one is always the state in which each counter track component is equal to $\cmtodaQuiescent$%
  ).
  We thus obtain
  that every trace of $\daAutomaton$ contains at most $(k+2)$ looping state.
  \qed
\end{proof}

\section{From distributed automata to counter machines}
\label{app:da-to-cm}

The purpose of this appendix
is to prove Proposition~\ref{prp:da-to-cm}
(in Section~\ref{sec:da-to-cm}),
which states that
quasi-acyclic distributed automata on \dipath{s}
can be simulated by copyless counter machines.
To this end,
we first make a precise analysis of the behavior
exhibited specifically by quasi-acyclic automata.

For any distributed automaton
$\daAutomaton = \tuple{\daStateSet,\daInitFunc,\daTransFunc,\daAcceptSet}$,
the transition function
$\daTransFunc \colon \daExtendedStateSet \times \daStateSet \to \daStateSet$
can be converted into a \defd{history transition function}
$\daHistoryTransFunc \colon
 \daExtendedStateSet^\omega \times \daStateSet \to \daStateSet^\omega$,
where $\daStateSet^\omega$ denotes the set of infinite sequences over $\daStateSet$.
This function takes as input
the infinite sequence of states traversed by some node~$\Node[1]$
and the initial state of $\Node[1]$'s successor~$\Node[2]$,
and returns the infinite sequence of states traversed by $\Node[2]$.
(If $\Node[2]$ has no predecessor,
the input sequence is simply~$\set{\daNoState}^\omega$.)
We can easily express $\daHistoryTransFunc$ recursively as follows,
for $\daState[1] \in \daExtendedStateSet$,\,
$\daStateSequence[1] \in \daExtendedStateSet^\omega$, and
$\daState[2] \in \daStateSet$:
\begin{equation*}
  \daHistoryTransFunc(\daState[1] \daStateSequence[1], \daState[2]) =
  \daState[2] \cdot
  \daHistoryTransFunc(\daStateSequence[1], \daTransFunc(\daState[1],\daState[2])).
\end{equation*}

Roughly speaking,
a sequential machine~$\cmMachine$ that simulates $\daAutomaton$
will evaluate $\daHistoryTransFunc$ instead of $\daTransFunc$.
While reading a word
$\Word = \Symbol_1 \dots \Symbol_n$
from left to right,
$\cmMachine$ keeps track of the infinite sequence of states
that $\daAutomaton$ would traverse
on the node corresponding to the currently read symbol~$\Symbol_i$.
So the transition function of $\cmMachine$
is basically an encoded version of $\daHistoryTransFunc$.
We will show that if $\daAutomaton$ is quasi-acyclic,
then $\cmMachine$ can evaluate~$\daHistoryTransFunc$
using only a finite-state memory and a fixed number of counters.
While this may be fairly obvious on an intuitive level,
the details are a bit cumbersome to formalize.

To keep our proof as clear as possible,
we divide it into three steps.
The first consists in the following lemma,
which describes the form of
$\daHistoryTransFunc(\daStateSequence[1], \daState[2])$
with respect to a given sequence of states $\daStateSequence[1]$.
This description is then reiterated
at a slightly higher level of abstraction
in Lemma~\ref{lem:encoded-transition},
and finally put to use
in the constructive proof of Proposition~\ref{prp:da-to-cm}.

\begin{lemma}
  \label{lem:sequence-transition}
  Let
  $\daAutomaton = \tuple{\daStateSet,\daInitFunc,\daTransFunc,\daAcceptSet}$
  be a quasi-acyclic distributed automaton of maximum trace length~$\daTraceLength$
  and $\daHistoryTransFunc$ be its history transition function.
  Given a state
  $\daState[2] \in \daStateSet$
  and an infinite, ultimately constant sequence
  \begin{equation*}
    \daStateSequence[1]
    = (\daState[1]_1)^{m_1} \cdots (\daState[1]_{r-1})^{m_{r-1}} \cdot (\daState[1]_r)^\omega
    \,\in\, \daExtendedStateSet^\omega,
  \end{equation*}
  where $m_1, \dots, m_{r-1} \in \Positive$,
  the derived sequence
  $\daHistoryTransFunc(\daStateSequence[1], \daState[2])$
  is of the form
  \begin{equation*}
    \daStateSequence[2]
    = \daState[2] \cdot
      \underbrace{\daState[2]_{1, 1} \cdots
                  \daState[2]_{1, d_1 - 1} \cdot
                  (\daState[2]_{1, d_1})^{n_1}
                 }_{\textstyle\text{stems from $(\daState[1]_1)^{m_1}$}}
      \cdots
      \underbrace{\daState[2]_{i, 1} \cdots
                  \daState[2]_{i, d_i - 1} \cdot
                  (\daState[2]_{i, d_i})^{n_i}
                 }_{\textstyle\text{stems from $(\daState[1]_i)^{m_i}$}}
      \cdots
      \underbrace{\daState[2]_{r, 1} \cdots
                  \daState[2]_{r, d_r - 1} \cdot
                  (\daState[2]_{r, d_r})^\omega
                 }_{\textstyle\text{stems from $(\daState[1]_r)^\omega$}},
  \end{equation*}
  such that the following properties are satisfied:
  \begin{enumerate}
  \item \label{itm:count}
    For $1 \leq i < r$,
    we have
    $1 \leq d_i \leq \min\set{\daTraceLength, m_i}$
    and $n_i = m_i - (d_i - 1)$;
    moreover,
    $1 \leq d_r \leq \daTraceLength$.
  \item \label{itm:distinct}
    The first~$d_i$ states
    $\daState[2]_{i, 1}, \dots, \daState[2]_{i, d_i}$
    that stem from $(\daState[1]_i)^{m_i}$,
    or from $(\daState[1]_r)^\omega$ for $i = r$,
    are pairwise distinct.
    Furthermore,
    if the very first state~$\daState[2]_{i, 1}$
    is equal to the state that immediately precedes it
    in~$\daStateSequence[2]$,
    then all the states stemming from
    $(\daState[1]_i)^{m_i}$
    are the same,
    \ie, $d_i = 1$.
  \item \label{itm:threshold}
    Any $m_i \geq \daTraceLength$
    can be replaced in $\daStateSequence[1]$
    by some other $m_i' \geq \daTraceLength$
    without affecting the form of $\daStateSequence[2]$.
    That is,
    we obtain the same values
    $d_1, \dots, d_r$
    and the same states
    $\daState[2], \daState[2]_{1, 1}, \dots, \daState[2]_{r, d_r}$.
    The only difference is
    that~$n_i$ gets replaced by
    $n_i' = m_i' - (d_i - 1)$.
  \end{enumerate}
\end{lemma}
\begin{proof}
  To see the form of~$\daStateSequence[2]$,
  we can simply “unroll” the recursive definition of~$\daHistoryTransFunc$
  and compute~$\daStateSequence[2]$ in an infinite number of steps.
  We initialize
  a variable~$\daStateSequence[2]_\curr$ to the empty sequence
  and
  another variable~$\daState[2]_\curr$ to the state~$\daState[2]$,
  and then let a third variable~$\daState[1]_\curr$ iterate
  over all states in~$\daStateSequence[1]$.
  In each step,
  we append~$\daState[2]_\curr$ to~$\daStateSequence[2]_\curr$
  and update~$\daState[2]_\curr$
  to $\daTransFunc(\daState[1]_\curr, \daState[2]_\curr)$.

  \emph{Statements~\ref{itm:count} and~\ref{itm:distinct}:}
  While we iterate over a subsequence of $\daStateSequence[1]$
  that has the form~$(\daState[1]_i)^{m_i}$,
  the value of~$\daState[1]_\curr$ remains the same.
  Hence,
  as soon as the value of~$\daState[2]_\curr$
  does not change between two successive iteration steps,
  it will remain the same
  for all the remaining steps over~$(\daState[1]_i)^{m_i}$.
  If $m_i > \daTraceLength$,
  then such a repetition of the same value must occur
  after at most~$\daTraceLength$ iterations,
  since the sequence of states traversed by~$\daState[2]_\curr$
  follows a trace in
  a quasi-acyclic automaton of maximum trace length~$\daTraceLength$.
  Therefore,
  the subsequence of~$\daStateSequence[2]$
  that stems from~$(\daState[1]_i)^{m_i}$
  must be of the form
  $\daState[2]_{i, 1} \cdots
   \daState[2]_{i, d_i - 1} \cdot
   (\daState[2]_{i, d_i})^{n_i}$,
  where
  the first~$d_i$ states are pairwise distinct,
  $1 \leq d_i \leq \min\set{\daTraceLength, m_i}$,
  and $n_i = m_i - (d_i - 1)$.
  By the same token,
  the subsequence
  that stems from~$(\daState[1]_r)^\omega$
  must be of the form
  $\daState[2]_{r, 1} \cdots
   \daState[2]_{r, d_r - 1} \cdot
   (\daState[2]_{r, d_r})^\omega$,
  where
  the first~$d_r$ states are pairwise distinct
  and
  $1 \leq d_r \leq \daTraceLength$.
  Moreover,
  if the first state~$\daState[2]_{i, 1}$
  of such a subsequence
  is the same as its immediate predecessor in~$\daStateSequence[2]$,
  then the mentioned repetition already occurs in the first iteration,
  and thus $d_i = 1$.

  \emph{Statement~\ref{itm:threshold}:}
  In view of the above,
  if $m_i \geq \daTraceLength$,
  then the actual value of~$m_i$ is irrelevant
  for determining the sequence
  $\daState[2]_{i, 1}, \dots,
   \daState[2]_{i, d_i - 1},
   \daState[2]_{i, d_i}$
  of distinct states stemming from~$(\daState[1]_i)^{m_i}$,
  since the length~$d_i$ is upper-bounded by~$\daTraceLength$.
  Hence,
  if we replace $m_i$ by some other $m_i' \geq \daTraceLength$,
  then iterating over $(\daState[1]_i)^{m_i'}$ will give rise to
  the subsequence
  $\daState[2]_{i, 1} \cdots
   \daState[2]_{i, d_i - 1} \cdot
   (\daState[2]_{i, d_i})^{n_i'}$,
  where $n_i' = m_i' - (d_i - 1)$.
  \qed
\end{proof}

In order to take advantage of Lemma~\ref{lem:sequence-transition},
we now have to encode an infinite sequence $\daStateSequence[1]$
of states traversed by $\daAutomaton$
in such a way that it fits into the memory of a counter machine.
This is easy if $\daAutomaton$ is quasi-acyclic,
because then $\daStateSequence[1]$ can always be represented
as a finite sequence of pairs
$\tuple{\daState[1]_1, m_1} \cdots \tuple{\daState[1]_{r-1}, m_{r-1}} \cdot \tuple{\daState[1]_r, \infty}$,
where
$\daState[1]_1, \dots, \daState[1]_r$
are pairwise distinct states of $\daAutomaton$,
$m_1, \dots, m_{r-1}$
are positive integers, and
$\infty$ is a symbol
that stands for an infinite number of repetitions.
Such a sequence is maximally compressed
in the sense that each state occurs in at most one pair.
Accordingly,
we define a \defd{compressed-history transition function}
\begin{equation*}
  \daCompressedHistoryTransFunc \colon
  \bigl( (\daExtendedStateSet \times \Positive)^\ast \times (\daExtendedStateSet \times \set{\infty}) \bigr)
  \times \daStateSet
  \,\to\,
  \bigl( (\daStateSet \times \Positive)^\ast \times (\daStateSet \times \set{\infty}) \bigr)
\end{equation*}
that operates on compressed sequences of states
in exactly the same way
as $\daHistoryTransFunc$ does on uncompressed ones.
The next lemma essentially restates Lemma~\ref{lem:sequence-transition}
in terms of compressed sequences and $\daCompressedHistoryTransFunc$.

\begin{lemma}
  \label{lem:encoded-transition}
  Let
  $\daAutomaton = \tuple{\daStateSet,\daInitFunc,\daTransFunc,\daAcceptSet}$
  be a quasi-acyclic distributed automaton of maximum trace length~$\daTraceLength$
  and $\daCompressedHistoryTransFunc$ be its compressed-history transition function.
  Given a state
  $\daState[2] \in \daStateSet$
  and a finite sequence of pairs
  \begin{equation*}
    \daCompressedStateSequence[1]
    = \tuple{\daState[1]_1, m_1} \cdots \tuple{\daState[1]_{r-1}, m_{r-1}} \cdot \tuple{\daState[1]_r, \infty}
    \,\in\, \bigl( (\daExtendedStateSet \times \Positive)^\ast \times (\daExtendedStateSet \times \set{\infty}) \bigr),
  \end{equation*}
  the derived sequence
  $\daCompressedHistoryTransFunc(\daCompressedStateSequence[1], \daState[2])$
  is of the form
  \begin{equation*}
    \daCompressedStateSequence[2]
    = \tuple{\daState[2]_1, n_1} \cdots \tuple{\daState[2]_{s-1}, n_{s-1}} \cdot \tuple{\daState[2]_s, \infty},
  \end{equation*}
  such that the following properties are satisfied:
  \begin{enumerate}
  \item \label{itm:encoded-count}
    Each number~$n_j$ can be expressed as either
    $n_j = 1$ or
    $n_j = m_{i_j} + \dots + m_{(i_j + k_j)} + c_j$,
    where $-\daTraceLength < c_j \leq 1$
    and each~$m_i$ occurs in the expression of at most one~$n_j$.
  \item \label{itm:encoded-threshold}
    Any $m_i \geq \daTraceLength$
    can be replaced in $\daCompressedStateSequence[1]$
    by some other $m_i' \geq \daTraceLength$
    without affecting the form of $\daCompressedStateSequence[2]$.
    That is,
    we obtain
    the same states
    $\daState[2]_1, \dots, \daState[2]_s$
    and the same expressions defining
    $n_1, \dots, n_{s-1}$
    with respect to
    $m_1, \dots, m_i', \dots, m_{r-1}$.
  \end{enumerate}
\end{lemma}
\begin{proof}
  Obviously,
  the infinite sequence of states represented by~$\daCompressedStateSequence[2]$
  is of the form
  \begin{equation*}
    \daStateSequence[2]'
    = (\daState[2]_1)^{n_1} \cdots (\daState[2]_{s-1})^{n_{s-1}} \cdot (\daState[2]_s)^\omega,
  \end{equation*}
  where
  $\daState[2]_1, \dots, \daState[2]_s$
  are pairwise distinct.
  By Lemma~\ref{lem:sequence-transition},
  we can also represent it as
  \begin{equation*}
    \daStateSequence[2]' =
    \daState[2] \cdot
    \underbrace{\daState[2]'_{1, 1} \cdots
                \daState[2]'_{1, d_1 - 1} \cdot
                (\daState[2]'_{1, d_1})^{n_1'}
               }_{\textstyle\text{stems from $\tuple{\daState[1]_1, m_1}$}}
    \cdots
    \underbrace{\daState[2]'_{i, 1} \cdots
                \daState[2]'_{i, d_i - 1} \cdot
                (\daState[2]'_{i, d_i})^{n_i'}
               }_{\textstyle\text{stems from $\tuple{\daState[1]_i, m_i}$}}
    \cdots
    \underbrace{\daState[2]'_{r, 1} \cdots
                \daState[2]'_{r, d_r - 1} \cdot
                (\daState[2]'_{r, d_r})^\omega
               }_{\textstyle\text{stems from $\tuple{\daState[1]_r, \infty}$}}.
  \end{equation*}
  Unlike the former representation,
  the latter is not maximally compressed
  but has the advantage of directly relating
  each occurrence of a state~$\daState[2]'_{i, j}$ in~$\daStateSequence[2]'$
  to the pair $\tuple{\daState[1]_i, m_i}$ in~$\daCompressedStateSequence[1]$
  that gives rise to it
  (through $\daCompressedHistoryTransFunc$).
  Note that
  simply-indexed states like $\daState[2]_j$
  refer to the former representation
  whereas
  doubly-indexed primed ones like $\daState[2]'_{i, j}$
  refer to the latter.

  Our goal is now to restate
  the findings of Lemma~\ref{lem:sequence-transition}
  in terms of the former representation.
  To this end,
  let us consider the sequence of indices $i_1, \dots, i_s$,
  where
  $i_1$ is equal to~$1$,
  and for $2 \leq j \leq s$,
  index $i_j$ identifies the pair
  $\tuple{\daState[1]_{i_j}, m_{i_j}}$
  in~$\daCompressedStateSequence[1]$ that gives rise to
  the first occurrence of~$\daState[2]_j$ in~$\daStateSequence[2]'$.
  Since a single pair $\tuple{\daState[1]_i, m_i}$
  can yield $d_i$ distinct states
  $\daState[2]'_{i, 1}, \dots, \daState[2]'_{i, d_i}$,
  it is possible for $i_j$ to be equal to $i_{j+1}$.

  \emph{Statement~\ref{itm:encoded-count}:}
  We first consider the case $j = 1$,
  where we must determine the number~$n_1$
  of occurrences of $\daState[2]_1 = \daState[2]$.
  If $\daState[2] \neq \daState[2]'_{1, 1}$,
  then $n_1 = 1$ and $i_2 = 1$,
  since this implies that $\daState[2]'_{1, 1}$
  is the first occurrence of $\daState[2]_2$.
  Otherwise (if $\daState[2] = \daState[2]'_{1, 1}$),
  Lemma~\ref{lem:sequence-transition}.\ref{itm:distinct}
  tells us that $d_1 = 1$,
  and thus by Lemma~\ref{lem:sequence-transition}.\ref{itm:count},
  the pair $\tuple{\daState[1]_1, m_1}$
  gives rise to $n_1' = m_1$ additional occurrences of $\daState[2]$.
  Similarly,
  if we also have $\daState[2]'_{1, d_1} = \daState[2]'_{2, 1}$,
  then $\tuple{\daState[1]_2, m_2}$
  gives rise to $n_2' = m_2$ further occurrences of~$\daState[2]$.
  This continues $k_1$ times,
  until we reach the first position $i_1 + k_1$
  such that $\daState[2]'_{(i_1+k_1), d_{i_1+k_1}} \neq \daState[2]'_{(i_1+k_1+1), 1}$.
  Hence,
  $n_1$ is of the form $m_{i_1} + \dots + m_{(i_1+k_1)} + 1$,
  and~$i_2$
  (the index of the pair in~$\daCompressedStateSequence[1]$
  that yields the first occurrence of~$\daState[2]_2$)
  is equal to $i_1 + k_1 + 1$.

  Next,
  we turn to the case $2 \leq j \leq s - 1$,
  which is very similar.
  If $\daState[2]_j$ is one of the first $d_{i_j} - 1$ states
  that stem from $\tuple{\daState[1]_{i_j}, m_{i_j}}$,
  \ie, one of
  $\daState[2]'_{i_j, 1}, \dots, \daState[2]'_{i_j, d_{i_j} - 1}$,
  then by Lemma~\ref{lem:sequence-transition}.\ref{itm:distinct}
  it is repeated only once in~$\daStateSequence[2]'$,
  and we have $n_j = 1$ and $i_{j+1} = i_j$.
  Otherwise,
  $\daState[2]_j = \daState[2]'_{i_j, d_{i_j}}$,
  which by Lemma~\ref{lem:sequence-transition}.\ref{itm:count}
  gives us $n_{i_j}' = m_{i_j} - (d_{i_j} - 1)$
  occurrences of $\daState[2]_j$.
  If additionally
  $\daState[2]'_{i_j, d_{i_j}} = \daState[2]'_{i_j + 1, 1}$,
  then
  by Lemma~\ref{lem:sequence-transition}.\ref{itm:distinct}
  and~\ref{lem:sequence-transition}.\ref{itm:count},
  we get another $n_{i_j + 1}' = m_{i_j + 1}$ occurrences of $\daState[2]_j$.
  This can be iterated $k_j$ times,
  until we reach the first position $i_j + k_j$
  such that
  $\daState[2]'_{(i_j+k_j), d_{i_j+k_j}} \neq \daState[2]'_{(i_j+k_j+1), 1}$.
  Consequently,
  $n_j$ is of the form $m_{i_j} + \dots + m_{(i_j + k_j)} - (d_{i_j} - 1)$,
  where $1 \leq d_{i_j} \leq \daTraceLength$
  (by Lemma~\ref{lem:sequence-transition}.\ref{itm:count}),
  and~$i_{j+1}$ is equal to $i_j + k_j + 1$.
  In conjunction with the previous paragraph,
  this also entails that
  each~$m_i$ occurs in the expression of at most one~$n_j$.

  For the sake of completeness,
  let us also mention the case $j = s$,
  which can be seen as analogous to the previous one.
  If we proceed as before,
  we get a value~$n_s$ of the form
  $m_{i_s} + \dots + m_{(i_s + k_s)} - (d_{i_s} - 1)$,
  where $i_s + k_s = r$.
  But since $m_r = \infty$,
  this expression can be simplified to~$\infty$,
  which yields the final pair $\tuple{\daState[2]_s, \infty}$
  in $\daCompressedStateSequence[2]$.

  \emph{Statement~\ref{itm:encoded-threshold}:}
  If $m_i \geq \daTraceLength$,
  then by Lemma~\ref{lem:sequence-transition}.\ref{itm:threshold},
  we can replace it in $\daCompressedStateSequence[1]$
  by some other $m_i' \geq \daTraceLength$
  without affecting the form of $\daStateSequence[2]'$.
  The only difference is that
  $n_i' = m_i - (d_i - 1)$ gets replaced by
  $n_i'' = m_i' - (d_i - 1)$.
  This does not affect the above description of $\daCompressedStateSequence[2]$,
  except that $m_i$ must be replaced by $m_i'$
  (if it occurs in the expression of some $n_j$).
  \qed
\end{proof}

We are now ready to prove the main proposition of Section~\ref{sec:da-to-cm},
since Lemma~\ref{lem:encoded-transition} provides us with the means
to encode the history transition function~$\daHistoryTransFunc$
of a quasi-acyclic distributed automaton~$\daAutomaton$
into the transition function
of a copyless counter machine~$\cmMachine$.

\begin{uProposition}[\ref{prp:da-to-cm}]
  For every quasi-acyclic distributed automaton
  with at most $(\daLoopNumMinusOne + 1)$ loops per trace
  and maximum trace length~$\daTraceLength$,
  we can effectively construct
  an equivalent copyless $\daLoopNumMinusOne$-counter machine
  with $\daTraceLength$-access.
\end{uProposition}
\begin{proof}
  Given a quasi-acyclic distributed automaton
  $\daAutomaton = \tuple{\daStateSet,\daInitFunc,\daTransFunc,\daAcceptSet}$
  over $\Alphabet$-labeled \dipath{s}
  with at most~$(\daLoopNumMinusOne+1)$ loops per trace
  and maximum trace length~$\daTraceLength$,
  we construct
  an equivalent copyless $\daLoopNumMinusOne$-counter machine~$\cmMachine$
  with $\daTraceLength$-access.
  Basically,
  after $\cmMachine$ has read
  the $i$-th symbol of the input word~$\Word$,
  its memory configuration will represent
  the temporal behavior exhibited by~$\daAutomaton$
  at the $i$-th node
  of the \dipath{} corresponding to~$\Word$.
  This exploits the quasi-acyclicity of~$\daAutomaton$
  to represent
  the infinite sequence of states traversed by a node
  as a finite sequence of pairs in
  $\daStateSet \times (\Positive \cup \set{\infty})$,
  where
  values other than $1$ and $\infty$ are stored in the counters.

  Formally,
  we define
  $\cmMachine = \tuple{\cmStateSet,\cmCounterSet,\cmInitState,\cmTransFunc,\cmAcceptSet}$,
  where
  \begin{align*}
    \cmStateSet &\subseteq
    \textstyle \bigcup_{1 \leq r \leq \daTraceLength}
    \Bigl(
    \bigl(\daExtendedStateSet \times (\cmCounterSet \cup \set{1})\bigr)^{r - 1}
    \times
    \bigl(\daExtendedStateSet \times \set{\infty}\bigr)
    \Bigr), \\
    \cmInitState &= \tuple{\daNoState,\infty}, \\
    \cmAcceptSet &=
    \bigsetbuilder{\cmState \in \cmStateSet}
               {\text{$\cmState$ contains a pair
                      $\tuple{\daState,\cmCounterConst} \in
                              \daAcceptSet \times (\cmCounterSet \cup \set{1,\infty})$}},
  \end{align*}
  and $\cmTransFunc$ is defined as follows.
  Given a state
  \begin{equation*}
    \cmState =
    \tuple{\daState[1]_1, \cmCounterConst_1}
    \cdots \tuple{\daState[1]_{r-1}, \cmCounterConst_{r-1}}
    \cdot \tuple{\daState[1]_r, \infty} \in \cmStateSet,
  \end{equation*}
  an $\daTraceLength$-truncated valuation
  $\cmTruncatedValuation =
   (\cut{-\daTraceLength}{+\daTraceLength} \circ \cmValuation) \in \range[-\daTraceLength]{\daTraceLength}^\cmCounterSet$,
  and a symbol $\Symbol \in \Alphabet$,
  we can determine
  the result $\tuple{\cmState', \cmUpdateFunc}$ of the transition
  $\cmTransFunc(\cmState, \cmTruncatedValuation, \Symbol)$
  in three steps.
  First,
  we consider the sequence
  \begin{equation*}
    \daCompressedStateSequence[1] =
    \tuple{\daState[1]_1, m_1}
    \cdots \tuple{\daState[1]_{r-1}, m_{r-1}}
    \cdot \tuple{\daState[1]_r, \infty}
  \end{equation*}
  that is obtained from $\cmState$
  by replacing each counter variable
  by the corresponding ($\daTraceLength$-truncated) value “seen” by~$\cmMachine$,
  \ie,
  $m_i = 1$ if $\cmCounterConst_i$ is the constant $1$,
  and
  $m_i = \cmTruncatedValuation(\cmCounterConst_i)$ if
  $\cmCounterConst_i$ is some counter variable in $\cmCounterSet$.
  Second,
  we obtain from~$\daCompressedStateSequence[1]$ the derived sequence
  \begin{equation*}
    \daCompressedStateSequence[2]
    = \daCompressedHistoryTransFunc(\daCompressedStateSequence[1], \daInitFunc(\Symbol))
    = \tuple{\daState[2]_1, n_1} \cdots \tuple{\daState[2]_{s-1}, n_{s-1}} \cdot \tuple{\daState[2]_s, \infty}.
  \end{equation*}
  (Since the values $m_1, \dots, m_{r-1}$
  are all bounded by $\daTraceLength$,
  this step can be precomputed and stored in a finite lookup table.)
  Third,
  we obtain the state~$\cmState'$ by replacing
  the number of occurrences~$n_j$ of each looping state~$\daState[2]_j$
  in~$\daCompressedStateSequence[2]$
  by some (arbitrarily chosen) counter variable~$\cmCounter_j$,
  such that every counter variable occurs at most once in~$\cmState'$.
  That is,
  we set
  \begin{equation*}
    \cmState' =
    \tuple{\daState[2]_1, \cmCounterConst'_1} \cdots
    \tuple{\daState[2]_{s-1}, \cmCounterConst'_{s-1}} \cdot
    \tuple{\daState[2]_s, \infty}
  \end{equation*}
  where
  $\cmCounterConst'_j = 1$ if $\daState[2]_j$ is a non-looping state,
  and
  $\cmCounterConst'_j = \cmCounter_j$ otherwise,
  such that $\cmCounter_j \neq \cmCounterConst'_i$ for $i \neq j$.
  Note that we have enough counter variables at our disposal
  because the number of looping states in
  $\daState[2]_1, \dots, \daState[2]_{s-1}$
  is at most~$\daLoopNumMinusOne$
  ($\daState[2]_s$ being necessarily a looping state).
  By Lemma~\ref{lem:encoded-transition}.\ref{itm:encoded-count},
  we know that
  each $n_j$ is either $1$ or of the form
  $m_{i_j} + \dots + m_{(i_j + k_j)} + c_j$,
  where the constant $c_j$ lies between $-\daTraceLength$ and $1$,
  and each~$m_i$ occurs in the expression of at most one~$n_j$.
  Hence,
  we define the update function~$\cmUpdateFunc$ such that
  for all $\cmCounter \in \cmCounterSet$,
  \begin{equation*}
    \cmUpdateFunc(\cmCounter) =
    \begin{cases*}
      \cmCounterConst_{i_j} + \dots + \cmCounterConst_{(i_j + k_j)} + c_j
      & if $\cmCounter = \cmCounterConst'_j$, \\
      0
      & otherwise.
    \end{cases*}
  \end{equation*}
  This function is copyless
  because each $\cmCounterConst_i$ is used at most once.
  Moreover,
  the total additive constant
  in each counter expression
  $\cmCounterConst_{i_j} + \dots + \cmCounterConst_{(i_j + k_j)} + c_j$
  lies between~$-\daTraceLength$ and~$\daTraceLength$,
  since
  we have $-\daTraceLength < c_j \leq 1$
  and
  there are at most
  $r - 1 \leq \daTraceLength - 1$
  terms~$\cmCounterConst_i$
  (which are either a counter variable or the constant~$1$).
  Therefore,
  $\cmUpdateFunc$ can be used
  in a counter machine with $\daTraceLength$-access.

  Now,
  consider any memory configuration
  $\tuple{\cmState, \cmValuation}$
  of~$\cmMachine$
  and any symbol~$\Symbol$ in~$\Alphabet$,
  and let
  $\tuple{\cmState', \cmValuation'}$
  be the corresponding successor memory configuration,
  \ie,
  $\cmTransFunc(\cmState, \cut{-\daTraceLength}{+\daTraceLength} \circ \cmValuation, \Symbol)
   = \tuple{\cmState', \cmUpdateFunc}$
  such that
  $\cmExtendedValuation \circ \cmUpdateFunc = \cmValuation'$.
  Furthermore,
  let~$\daCompressedStateSequence[1]$ and~$\daCompressedStateSequence[2]$
  be the sequences of pairs in
  $\daStateSet \times (\Positive \cup \set{\infty})$
  represented by~$\tuple{\cmState, \cmValuation}$
  and~$\tuple{\cmState', \cmValuation'}$,
  respectively.
  If~$\cmValuation$ assigns
  values of at most~$\daTraceLength$
  to all counters,
  \ie, if $\cut{-\daTraceLength}{+\daTraceLength} \circ \cmValuation = \cmValuation$,
  then we know by construction that
  $\daCompressedStateSequence[2]
   = \daCompressedHistoryTransFunc(\daCompressedStateSequence[1], \daInitFunc(\Symbol))$.
  Otherwise,
  Lemma~\ref{lem:encoded-transition}.\ref{itm:encoded-threshold} tells us that
  the sequence of states in~$\daCompressedStateSequence[2]$
  and the expressions defining each number of occurrences
  remain the same
  if we cut off the counter values at $\daTraceLength$.
  Consequently,
  we also have
  $\daCompressedStateSequence[2]
   = \daCompressedHistoryTransFunc(\daCompressedStateSequence[1], \daInitFunc(\Symbol))$.
  \qed
\end{proof}

\end{document}